\documentclass[leqno]{amsart}

\usepackage[foot]{amsaddr}

\usepackage{color}
\usepackage{geometry}
\usepackage{subfiles}
\usepackage{graphicx}
\usepackage[section]{placeins}
\usepackage{hyperref}
\usepackage{stmaryrd}
\usepackage{amsmath,amssymb,amsfonts,amsthm,textcomp}
\usepackage{mathtools}
\usepackage{tensor}
\usepackage{multicol}
\usepackage{subfig}
\usepackage{enumitem}

\newfont{\tenbfsl}{cmbxti9 scaled 1200}
\newfont{\tenbbb}{msbm10}
\newfont{\svnbbb}{msbm8}

\newcommand{\bs}[1]{\boldsymbol{#1}}
\newcommand{\cl}[1]{\mathcal{#1}}
\newcommand{\mt}[1]{\mathit{#1}}

\newcommand{\id}{\mbox{\tenbfsl 1\/}}

\newcommand{\fr}[2]{\textstyle{\frac{{#1}}{{#2}}}}

\newcommand{\dv}{\,\mathrm{d}v}
\newcommand{\da}{\,\mathrm{d}a}
\newcommand{\ds}{\,\mathrm{d}s}

\newcommand{\bdy}{\cl{B}}
\newcommand{\dbdy}{\partial\cl{B}}
\newcommand{\prt}{\cl{P}}
\newcommand{\dprt}{\partial\cl{P}}
\newcommand{\srf}{\cl{S}}
\newcommand{\dsrf}{\partial\cl{S}}
\newcommand{\edg}{\cl{C}}

\newcommand{\trans}{\scriptscriptstyle\mskip-1mu\top\mskip-2mu}

\newcommand{\tr}{\mathrm{tr}\mskip2mu}
\newcommand{\sym}{\mathrm{sym}\mskip2mu}
\newcommand{\skw}{\mathrm{skw}\mskip2mu}

\newcommand{\Grad}{\mathrm{grad}\mskip2mu}
\newcommand{\Div}{\mathrm{div}\mskip2mu}

\newcommand{\Grads}{\Grad_{\mskip-2mu\scriptscriptstyle\cl{S}}}
\newcommand{\Divs}{\Div_{\mskip-6mu\scriptscriptstyle\cl{S}}}

\newcommand{\pards}[2]{\frac{\partial{#1}}{\partial{#2}}}
\newcommand{\lpards}[2]{\partial{#1}/\partial{#2}}
\newcommand{\dd}[2]{\frac{\mathrm{d}{#1}}{\mathrm{d}{#2}}}

\newcommand{\surp}[1]{\{\!\!\{ {#1} \}\!\!\}}

\newtheorem{thm}{Theorem}

\newtheorem{prop}{Proposition}
\theoremstyle{remark}
\newtheorem{rmk}{Remark}
\theoremstyle{definition}

\newcommand{\xis}{\xi_{\scriptscriptstyle\cl{S}}}
\newcommand{\sigmas}{{\sigma_{\mskip-1mu\scriptscriptstyle\cl{S}}}}
\newcommand{\tauds}{\tau_{\scriptscriptstyle\partial\cl{S}}}
\newcommand{\tauc}{\tau_{\scriptscriptstyle\cl{C}}}
\newcommand{\bsvarpis}{\bs{\varpi}_{\mskip-2mu\scriptscriptstyle\cl{S}}}

\begin{document}

\title{Phase-field gradient theory}
\author{Luis Espath$^\sharp$ \& Victor Calo$^\flat$}
\address{$^\sharp$Department of Mathematics, RWTH Aachen University, Kackertstra{\ss}e 9 C363, 52072 Aachen, Germany.}
\email{$^\sharp$espath@gmail.com}
\address{$^\flat$School of Earth and Planetary Sciences, Curtin University, Kent Street, Bentley, Perth, WA 6102, Australia.}
\address{\phantom{$^\flat$}Curtin Institute for Computation, Curtin University, Kent Street, Bentley, Perth, WA 6102, Australia.}
\address{\phantom{$^\flat$}Mineral Resources, Commonwealth Scientific and Industrial Research Organisation (CSIRO), 10 Kensington, Perth, WA 6152, Australia.}
\email{$^\flat$vmcalo@gmail.com}

\date{\today}

\begin{abstract}
\noindent
We propose a phase-field theory for enriched continua. To generalize classical phase-field models, we derive the \emph{phase-field gradient theory} based on balances of microforces, microtorques, and mass. We focus on materials where second gradients of the phase field describe long-range interactions. By considering a nontrivial interaction inside the body, described by a boundary-edge microtraction, we characterize the existence of a microhypertraction field, a central aspect of this theory. On surfaces, we define the surface microtraction and the surface-couple microtraction emerging from internal surface interactions. We explicitly account for the lack of smoothness along a curve on surfaces enclosing arbitrary parts of the domain. In these rough areas, internal-edge microtractions appear. We begin our theory by characterizing these tractions. Next, in balancing microforces and microtorques, we arrive at the field equations. Subject to thermodynamic constraints, we develop a general set of constitutive relations for a phase-field model where its free-energy density depends on second gradients of the phase field. A priori, the balance equations are general and independent of constitutive equations, where the thermodynamics constrain the constitutive relations through the free-energy imbalance. To exemplify the usefulness of our theory, we generalize two commonly used phase-field equations. We propose a `generalized Swift--Hohenberg equation'---a second-grade phase-field equation---and its conserved version, the `generalized phase-field crystal equation'---a conserved second-grade phase-field equation. Furthermore, we derive the configurational fields arising in this theory. We conclude with the presentation of a comprehensive, thermodynamically consistent set of boundary conditions.\\
\textbf{AMS subject classifications:}
$\cdot$
74N20 
$\cdot$
80A22 
$\cdot$
80A17 
$\cdot$
82C26 
$\cdot$
35L65 
$\cdot$

\end{abstract}

\maketitle

\tableofcontents                        


\section{Introduction}
\label{introduction}

In this work, we propose a phase-field theory for enriched continua. This continuum framework extends and complements the work by Fried \& Gurtin \cite{Fri93,Fri94} and Gurtin \cite{Gur96}. However, we base our theory on Fosdick's approach \cite{Fos89,Fos16} to derive it. We generalize the Swift--Hohenberg equation \cite{Swi77}---the second-grade phase-field equation---and its conserved version---the phase-field crystal equation. We build our theory on balances of microforces and microtorques while accounting for `rough' arbitrary parts. Additionally, we present the configurational fields arising in this theory with its balance.

Brazovski\v{\i} \cite{Bra75} introduced the free-energy functional that delivers the equations of Swift--Hohenberg and phase-field-crystal. In the literature, these equations are typically found in the following form of gradient flows
\begin{equation}
\dot{\varphi}=-\dfrac{\delta\Psi}{\delta\varphi}\qquad\text{and}\qquad\dot{\varphi}=\Div\Grad\left(\dfrac{\delta\Psi}{\delta\varphi}\right),
\end{equation}
where $\varphi$ is the phase field (left: nonconserved case, right: conserved case) and $\Psi$ a free-energy functional depending on $(\varphi,\Grad\varphi,\Grad^2\varphi)$.

The outline of the work is as follows. In \S\ref{field.equations}, to derive this continuum theory, we allow for different types of interactions between adjacent arbitrary parts $\prt$ occurring inside the body $\bdy$ to characterize the primitive and fundamental contact microforce fields. Parts $\prt$ of the body $\bdy$ are arbitrary and may exhibit a lack of smoothness on their boundaries $\dprt$ along a curve $\edg\subset\dprt$. Thus, by balancing these traction fields, we arrive at the field equations of the \emph{phase-field gradient theory}. In \S\ref{power.statement}, we present the virtual power theorem. In \S\ref{thermodynamics}, we derive the thermodynamic laws, in the form of the energy balance and the free-energy imbalance to derive suitable constitutive equations. In \S\ref{nonconserved.case}, we introduce the nonconserved second-grade phase-field equation, its specialization to the Swif--Hohenberg equation, and the configurational fields with its balance equation. In \S\ref{conserved.case}, we aim at the conserved second-grade phase-field equation by augmenting the nonconserved version with a balance of mass and emulate the developments of \S\ref{nonconserved.case}. In \S\ref{boundary.conditions}, we derive thermodynamically consistent boundary conditions for both, the nonconserved and conserved cases. In \S\ref{conclusions}, we summarize this work. In appendix \S\ref{identities}, we present the mathematical identities used to derive this theory.

\section{Fundamental fields \& field equations}
\label{field.equations}

The continuum theory by Fried \& Gurtin \cite{Fri93,Fri94} and Gurtin \cite{Gur96} represents a turning point in phase-field theories. In this collection of papers, Fried \& Gurtin describe these phenomena from a mechanistic standpoint and derived the `generalized Allen--Cahn (Ginzburg--Landau)' and the `generalized Cahn--Hilliard' equations by introducing a balance of microforces. Their work makes explicit the underlying ‘forces' that dictate the evolution of phase fields. Herein, we denote these equations as first-grade phase-field descriptions. In this section, we characterize the fundamental fields, based on Fosdick's approach \cite{Fos16}, and derive the field equations that yield the \emph{phase-field gradient theory}.

\subsection{Fundamental fields}

\begin{figure}[!htb]
\centering
  \includegraphics[width=0.475\textwidth]{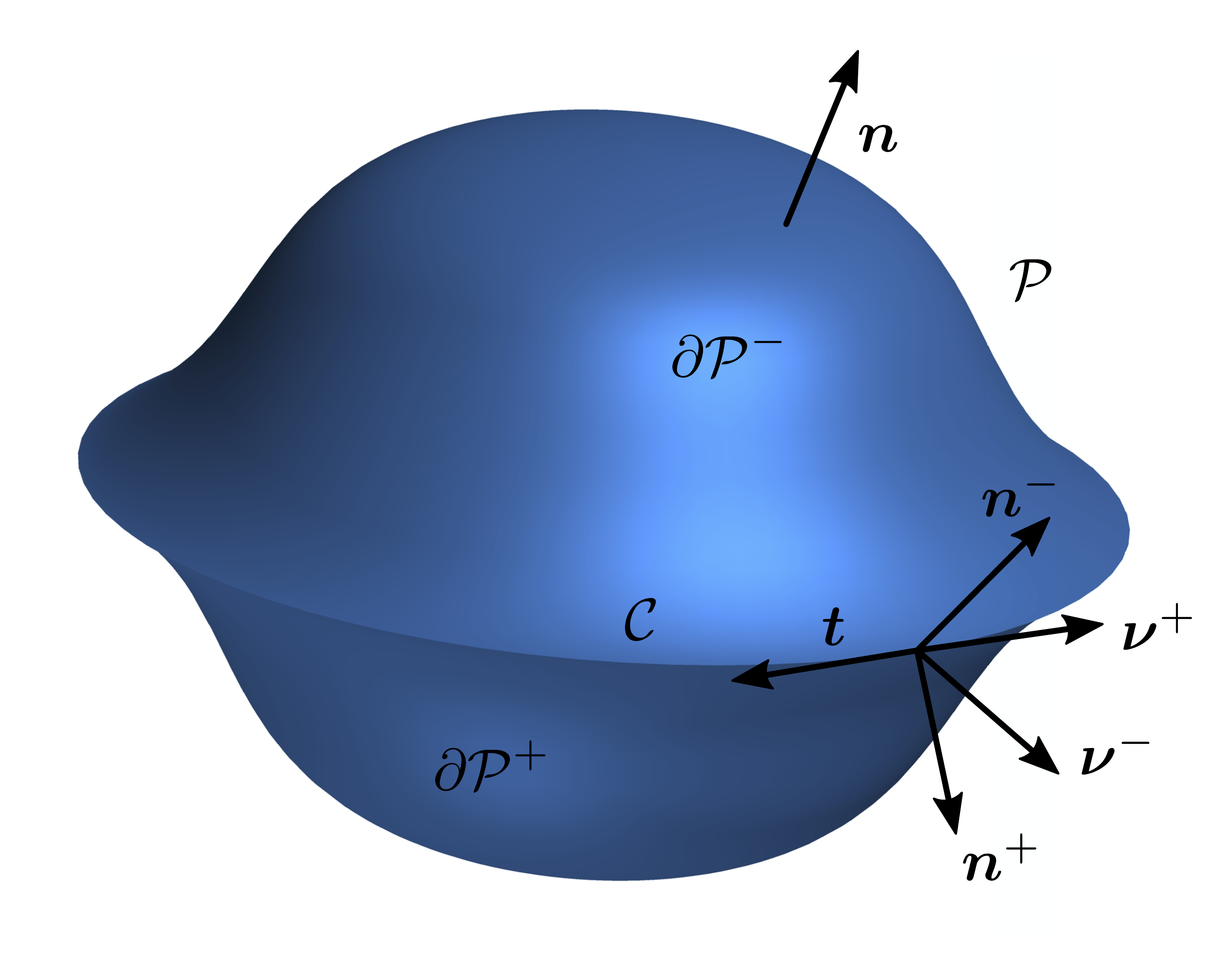}
  \caption{Arbitrary fixed part $\prt\subseteq\bdy$ with closed boundary surface $\dprt$ oriented by an outward unit normal $\bs{n}$. The boundary surface $\dprt$ lacks of smoothness along an internal-edge $\edg$. The pair $\{\bs{n}^+,\bs{n}^-\}$ characterizes the internal-edge $\edg$ which in turn is oriented by the unit tangent $\bs{t}$, where $\bs{t}\coloneqq\bs{n}^+\times\bs{\nu}^+=-\bs{n}^-\times\bs{\nu}^-$.}
\label{fg:part}
\end{figure}

Throughout what follows, $\bdy$ denotes a fixed region of a three-dimensional point space $\cl{E}$. $\prt\subseteq\bdy$ is an arbitrarily fixed subregion of $\bdy$ with a closed boundary surface $\dprt$ oriented by an outward unit normal $\bs{n}$. The surface $\dprt$ can lose its smoothness along a curve, namely an internal-edge $\edg$. Analyzing a neighborhood of an internal-edge $\edg$, two smooth surfaces $\dprt^{\pm}$ are defined. Thus, the limiting unit normals of $\dprt^{\pm}$ at $\edg$ are denoted by the pair $\{\bs{n}^+,\bs{n}^-\}$, which characterizes the internal-edge $\edg$. Equivalently, the limiting outward unit tangent-normal of $\dprt^{\pm}$ at $\edg$ are $\{\bs{\nu}^+,\bs{\nu}^-\}$. As a notational agreement, $\edg$ is oriented by the unit tangent $\bs{t}\coloneqq\bs{t}^+$ such that $\bs{\nu}^+\coloneqq\bs{t}^+\times\bs{n}^+$. Irrespectively of the surface $\srf$ being a boundary of $\prt$ the internal-edge remains a feature of the surface and not part of $\prt$. Furthermore, the body $\bdy$ and all its parts are open sets in $\cl{E}$. Figure \ref{fg:part} depicts the part under discussion.

\begin{figure}[!htb]
\centering
  \begingroup
  \captionsetup[subfigure]{width=0.475\textwidth}
  \subfloat[Smooth open surface $\srf$ oriented by an unit normal $\bs{n}$ with a boundary-edge $\dsrf$ oriented by a tangent unit $\bs{t}$.]{\label{fg:1}\includegraphics[width=0.475\textwidth]{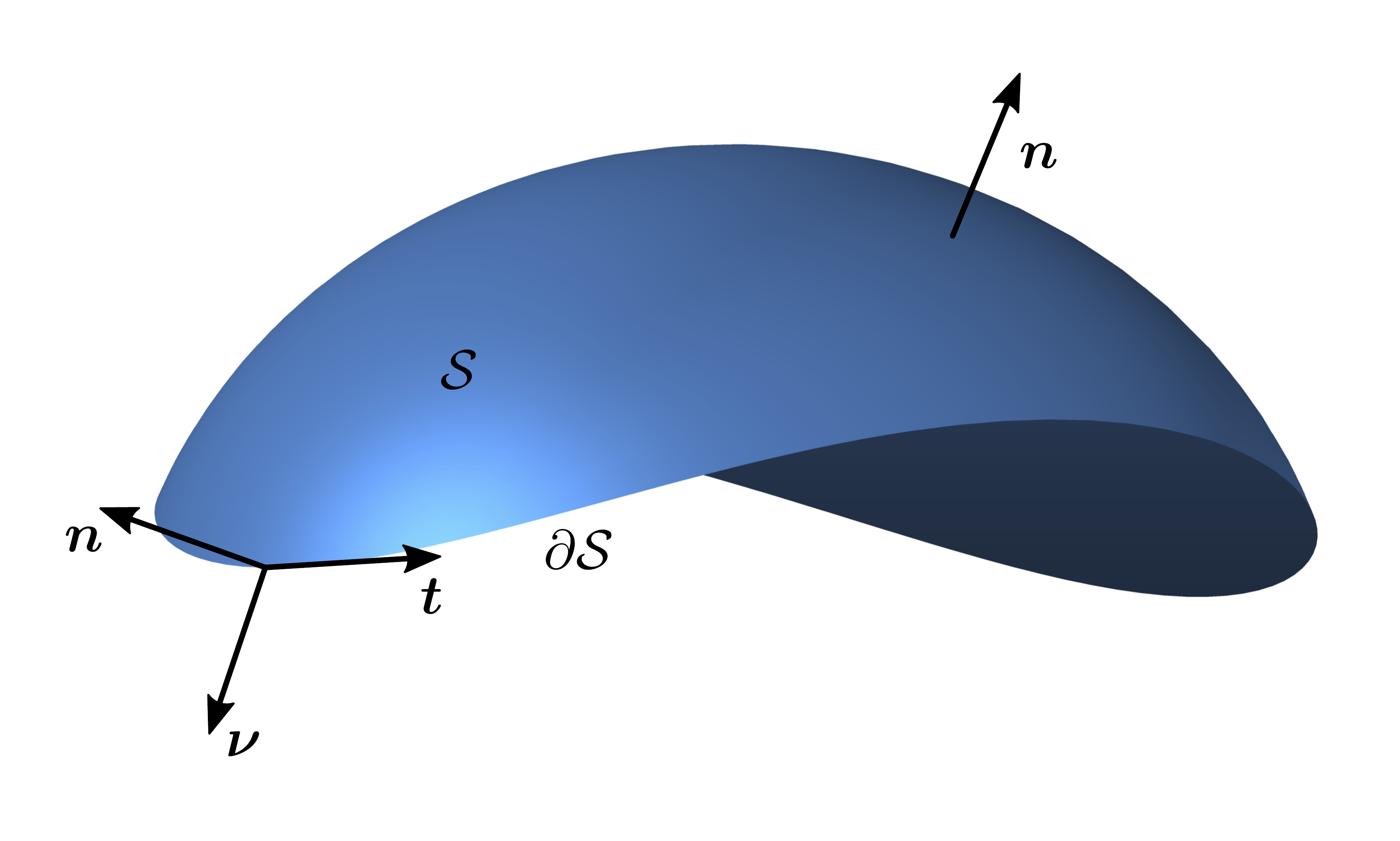}} \hspace{0.5cm}
  \subfloat[Nonsmooth open surface $\srf$ oriented by an unit normal $\bs{n}$ with a boundary-edge $\dsrf$ oriented by a tangent unit $\bs{t}$ and an internal-edge $\edg$ defined by the unit normals $\{\bs{n}^+,\bs{n}^-\}$ and oriented by the unit tangent $\bs{t}\coloneqq\bs{t}^+$.]{\label{fg:2}\includegraphics[width=0.475\textwidth]{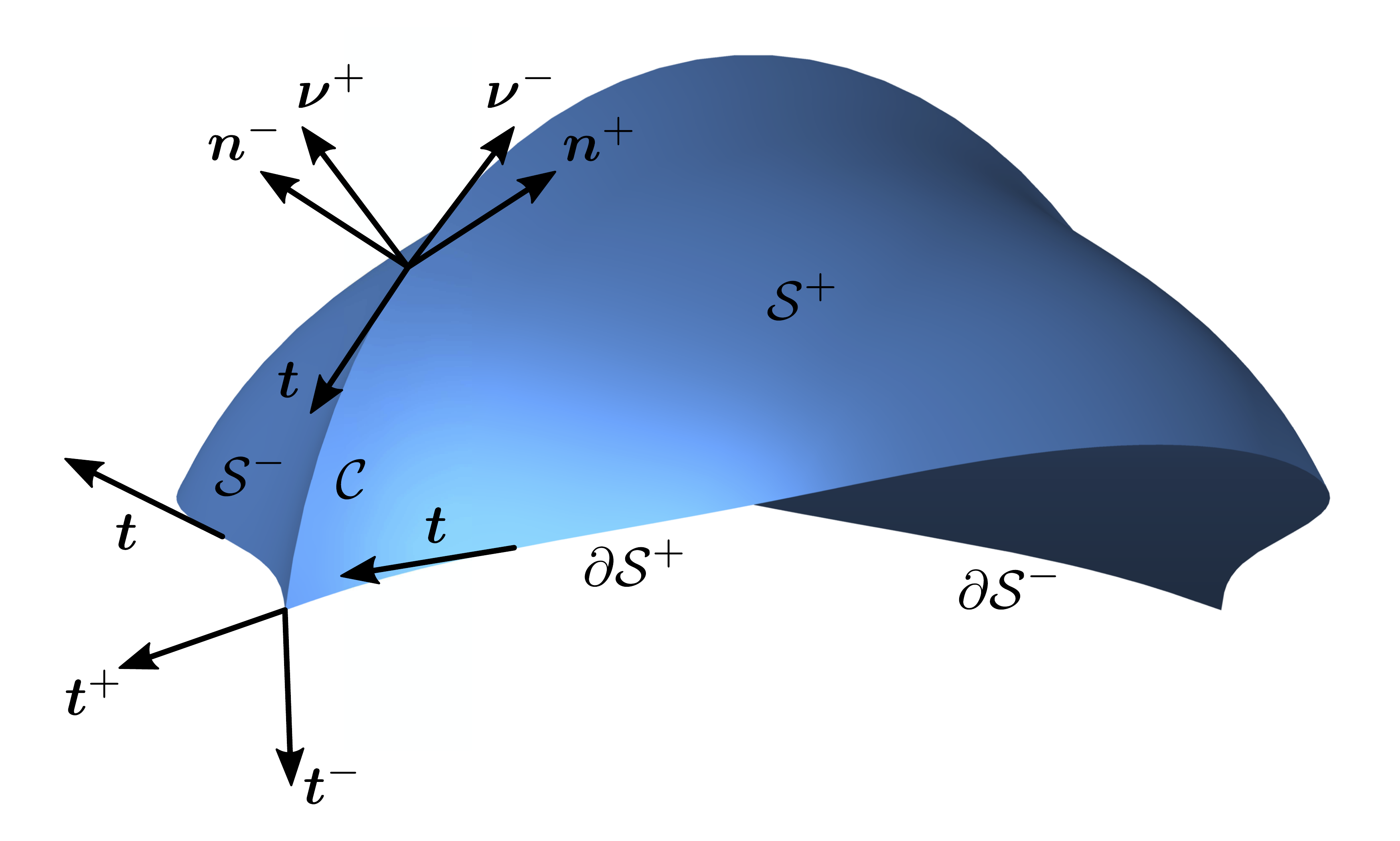}}
  \endgroup
  \caption{Surfaces.}
\label{fg:surfaces}
\end{figure}

To unfold the implications of considering arbitrary parts $\prt$ that lack of smoothness at an internal-edge $\edg$ arising in this theory, we begin by discussing the interaction of a smooth open surface $\srf\subseteq\prt$ with a boundary-edge $\dsrf$ and its adjacent parts of $\prt$, that is $\prt\setminus\srf$. Here, $\srf$ is oriented by an unit normal $\bs{n}$ with a boundary-edge $\dsrf$ oriented by the unit tangent $\bs{t}$. Boundary-edges $\dsrf$ are equipped with an intrinsic Darboux frame, composed by the unit tangent $\bs{t}$, unit normal $\bs{n}$, and outward unit tangent-normal $\bs{\nu}$, where $\bs{\nu}\coloneqq\bs{t}\times\bs{n}$. Understanding the underlying mechanical interactions of a boundary-edge $\dsrf$ allows us to further the understanding of an internal-edge $\edg$. Figure \ref{fg:surfaces} depicts a smooth open surface (left) and a nonsmooth open surface (right), which will serve to study boundary- and internal-edge interactions.

Before developing the field equations of the \emph{phase-field gradient theory}, we describe the set of interactions of adjacent parts of $\bdy$ and the resulting fundamental fields as follows.\\[6pt]

\paragraph{\emph{(i) Surface microtraction}:}

The surface microtraction $\xis\coloneqq\xis(\bs{x},t;\bs{n},\bs{L})$ represents a microforce per unit area, acting on an oriented surface $\srf\subset\bdy$ at $\bs{x}\in\srf$. It depends on $\srf$ through the pair $\{\bs{n},\bs{L}\}$, its outward unit normal $\bs{n}$ and its curvature tensor $\bs{L}$ (or the negative surface gradient of the unit normal, $-\Grads\bs{n}$). On the opposite side of $\srf$, $\srf^\ast$, the surface microtraction $\xis^\ast\coloneqq\xis^\ast(\bs{x},t;\bs{n},\bs{L})$ is defined. We say that $\xis^\ast$ is the intrinsic counterpart of $\xis$. Each of these surface microtractions are developed by the contact of each side of a surface $\srf$ with the adjacent parts of $\bdy$.\\[6pt]

\paragraph{\emph{(ii) Surface-couple microtraction}:}

The surface-couple microtraction $\bsvarpis\coloneqq\bsvarpis(\bs{x},t;\bs{n})$ represents a microtorque per unit area, acting on an oriented surface $\srf\subset\bdy$ at $\bs{x}\in\srf$. It depends on $\srf$ through its outward unit normal $\bs{n}$. On the opposite side of $\srf$, $\srf^\ast$, the surface-couple microtraction $\bsvarpis^\ast\coloneqq\bsvarpis^\ast(\bs{x},t;\bs{n})$ is defined. We say that $\bsvarpis^\ast$ is the intrinsic counterpart of $\bsvarpis$. Each of these surface-couple microtractions are also developed by the contact of each side of a surface $\srf$ with the adjacent parts of $\bdy$.\\[6pt]

\paragraph{\emph{(ii) Boundary-edge microtraction}:}

The boundary-edge microtraction $\tauds\coloneqq\tauds(\bs{x},t;\bs{\nu},\bs{n})$ represents a microforce per unit length, acting on an boundary-edge $\dsrf$ of an open oriented surface $\srf\subset\bdy$ at $\bs{x}\in\dsrf$. It depends on $\srf$ through the pair $\{\bs{\nu},\bs{n}\}$, its outward unit tangent-normal $\bs{\nu}$ and its unit normal $\bs{n}$. The boundary-edge microtraction is developed by the contact of a boundary-edge $\dsrf$ with the adjacent parts of $\bdy\setminus\srf$. The analysis of the boundary-edge microtraction and its definition is intrinsic to the study of open oriented surfaces. Thus, the boundary-edge microtraction does not contribute to balances that are reckoned on arbitrary parts $\prt$. However, the boundary-edge microtraction characterizes the internal-edge microtraction and more importantly, it supports the existence of a hypermicrostress field.\\[6pt]

\paragraph{\emph{(iv) Internal-edge microtraction}:}

The internal-edge microtraction $\tauc\coloneqq\tauc(\bs{x},t;\bs{n}^+,\bs{n}^-)$ represents a microforce per unit length, acting on an internal-edge $\edg$ of a nonsmooth oriented surface $\srf\subset\bdy$ at $\bs{x}\in\edg\subset\srf$. It depends on $\srf$ through the pair $\{\bs{n}^+,\bs{n}^-\}$, its unit normals defined at each smooth part of $\srf$. The internal-edge microtraction is developed by the contact of an internal-edge $\edg$ with the adjacent parts of $\bdy\setminus\srf$.

\paragraph{\emph{(v) External microforce}:}

The external microforce $\gamma\coloneqq\gamma(\bs{x},t)$ represents a body microforce per unit mass, acting on the body $\bdy$. It is developed by external causes, outside of $\bdy$.\\[6pt]

In accounting for the surface-couple, the boundary-edge, and the internal-edge microtractions, in addition to the conventional surface microtraction and the external microforce, we obtain a general second-grade phase-field theory. Conversely, first-grade phase-field theories introduced by Fried \& Gurtin \cite{Fri93,Fri94} and Gurtin \cite{Gur96} are recovered when these additional microtraction fields are neglected.

\subsection{Differential relations on evolving surfaces}

We began our theory by assuming the body $\bdy$ to be fixed, and so any other part $\prt$ and surface $\srf$. However, let the surfaces depicted in Figure \ref{fg:surfaces} undergo deformation from $\srf$ to $\bs{y}(\srf)\eqqcolon\srf_\epsilon$. This motion is somewhat arbitrary but smooth and needed to characterize variationally the contact microforce and microtorque fields arising from the interactions between adjacent parts. We then parameterize the motion from $\srf$ to $\srf_\epsilon$, with a one-parameter family of smooth invertible mappings, such that
\begin{equation}
\bs{y}_\epsilon\coloneqq\bs{x}+\epsilon\bs{u}(\bs{x}),\quad\epsilon\in[0,1],
\end{equation}
with $\bs{x}\in\srf$, $\bs{y}_\epsilon|_{\epsilon=0}=\bs{x}$, $\bs{y}_\epsilon|_{\epsilon=1}=\bs{y}$, $\bs{y}_\epsilon(\srf)\eqqcolon\srf_\epsilon\subset\bdy$, $\forall\,\epsilon\in[0,1]$, and the displacement $\bs{u}$ sufficiently smooth. Quantities with the subscript $\epsilon$ live on $\srf_\epsilon$.

Defining the parameterized deformation gradient of this motion by
\begin{equation}
\bs{F}\coloneqq\Grad\bs{y}_\epsilon\vert_{\epsilon=1}\qquad\text{and}\qquad\bs{F}_\epsilon\coloneqq\Grad\bs{y}_\epsilon,
\end{equation}
the unit tangent and unit normal at $\bs{y}_\epsilon$ on $\srf_\epsilon$, respectively, are
\begin{equation}
\bs{t}_\epsilon=\dfrac{1}{|\bs{F}_\epsilon\bs{t}|}\bs{F}_\epsilon\bs{t}\qquad\text{and}\qquad\bs{n}_\epsilon=\dfrac{1}{|\bs{F}^{-\trans}_\epsilon\bs{n}|}\bs{F}^{-\trans}_\epsilon\bs{n},
\end{equation}
where $(\cdot)^{\trans}$ represents the transposition. The curvature tensor $\bs{L}$ is
\begin{equation}\label{eq:curvature.tensor.L}
\bs{L}\coloneqq-\Grads\bs{n}=-(\Grad\bs{n})\bs{P}\qquad\text{and}\qquad\bs{P}\coloneqq\id-\bs{n}\otimes\bs{n},
\end{equation}
and the determinant of the deformation gradient $J\coloneqq\mathrm{det}(\bs{F})$. In a Darboux frame, the following relations hold,
\begin{spreadlines}{0.75em}
\begin{subequations}\label{eq:identities}
\begin{align}
\dd{\bs{t}_\epsilon}{\epsilon}\Big\vert_{\epsilon=0}&=\pards{\bs{u}}{s}-\left(\pards{\bs{u}}{s}\cdot\bs{t}\right)\bs{t},\\
\dd{\bs{n}_\epsilon}{\epsilon}\Big\vert_{\epsilon=0}&=-(\Grads\bs{u})^{\trans}\bs{n},\\
\dd{\bs{\nu}_\epsilon}{\epsilon}\Big\vert_{\epsilon=0}&=-\left(\pards{\bs{u}}{s}\cdot\bs{\nu}\right)\bs{t}+(\Grads\bs{u})^{\trans}\bs{n}\times\bs{t},\\
\dd{\bs{L}_\epsilon}{\epsilon}\Big\vert_\epsilon&=\Grads\left((\Grads\bs{u})^{\trans}\bs{n}\right)-\bs{L}\mskip+2.5mu\Grads\bs{u}+\bs{L}(\Grads\bs{u})^{\trans}\bs{n}\otimes\bs{n},\\
\dd{|\bs{F}_\epsilon\bs{t}|}{\epsilon}\Big\vert_{\epsilon=0}&=\Grad\bs{u}\colon\bs{t}\otimes\bs{t},\\
\dd{(|\bs{F}^{-\trans}_\epsilon\bs{n}|)}{\epsilon}\Big\vert_{\epsilon=0}&=-\Grad\bs{u}\colon\bs{n}\otimes\bs{n},\\
\dd{(J|\bs{F}^{-\trans}_\epsilon\bs{n}|)}{\epsilon}\Big\vert_{\epsilon=0}&=\Divs\bs{u}.
\end{align}
\end{subequations}
\end{spreadlines}
In appendix \ref{identities}, we detail the derivations leading to the relations \eqref{eq:identities}, which we use in this section. With these identities, we have suitable machinery to characterize variationally all types of microtractions we introduce in our description.

\subsection{Analysis on a smooth open oriented surface $\srf$}

To characterize boundary-edge microtractions, we postulate that the following surface balance of microforces holds on a smooth open oriented surface $\srf\in\bdy$,
\begin{equation}\label{eq:functional.H}
\cl{H}(\srf)\coloneqq\int\limits_{\srf}({\xis}+{\xis^\ast})\da+\int\limits_{\dsrf}{\tauds}\ds=0, \quad\forall\,\srf\subset\bdy\quad\text{and}\quad t,
\end{equation}
where the functional $\cl{H}$ on $\srf_\epsilon$ specializes to
\begin{equation}\label{eq:functional.H.epsilon}
\cl{H}(\srf_\epsilon)=\int\limits_{\srf_\epsilon}({\xis}_\epsilon+{\xis^\ast}_\epsilon)\da_\epsilon+\int\limits_{\dsrf_\epsilon}{\tauds}_\epsilon\ds_\epsilon.
\end{equation}

The first variation of $\cl{H}$ on $\srf$, which in turn is linear on $\bs{u}$, will serve us to characterize the boundary-edge microtraction and, consequently, will provide evidence of the existence of a hypermicrostress field. Thus, in what follows we analyze
\begin{equation}
\delta\cl{H}(\srf)[\bs{u}]=0,
\end{equation}
where
\begin{equation}
\delta\cl{H}(\srf)[\bs{u}]\coloneqq\dd{}{\epsilon}\cl{H}(\srf_\epsilon)|_{\epsilon=0},\quad\forall\,\,\mathrm{smooth}\,\,\bs{u}(\bs{x})\quad\text{and}\quad\forall\,\srf\subset\bdy.
\end{equation}

\subsubsection{Boundary-edge microtraction characterization and its implications: part 1}

In analyzing the second integral in \eqref{eq:functional.H.epsilon}, let
\begin{equation}
\cl{F}(\srf_\epsilon)\coloneqq\int\limits_{\dsrf_\epsilon}{\tauds}_\epsilon\ds_\epsilon=\int\limits_{\dsrf}\tauds(\bs{y}_\epsilon,t;\bs{\nu}_\epsilon,\bs{n}_\epsilon)|\bs{F}_\epsilon\bs{t}|\ds,
\end{equation}
be a functional on $\srf_\epsilon$. Additionally, let ${\tauds}_{\bs{x}}$, ${\tauds}_{\bs{\nu}}$, and ${\tauds}_{\bs{n}}$ be the intrinsic partial derivatives of $\tauds$ with respect to $\bs{x}$, $\bs{\nu}$, and $\bs{n}$, respectively. The partial derivative ${\tauds}_{\bs{x}}$ need not any additional characterization. The remainder derivatives, ${\tauds}_{\bs{\nu}}$ and ${\tauds}_{\bs{n}}$ need further understanding as follows.

\begin{rmk}[\textbf{\textit{Characterization of the partial intrinsic derivatives of the boundary-edge microtraction}}]\label{rmk:intrinsic.derivatives.edge}
\sloppy Let $\bs{\alpha}(\epsilon)$ and $\bs{\beta}(\epsilon)$ be smooth parametric orthonormal-vector-valued functions of $\epsilon\in[0,1]$, that is, $\bs{\alpha}(\epsilon),\bs{\beta}(\epsilon)\colon[0,1]\mapsto\mathrm{Unit}$\footnote{`$\mathrm{Unit}$' is a vector space $\cl{V}$ where all elements $\bs{v}$ have unit norm.} and $\bs{\alpha}(\epsilon)\cdot\bs{\beta}(\epsilon)=0$ $\forall\,\epsilon\in[0,1]$, where $\bs{\alpha}(0)\coloneqq\bs{\nu}$ and $\bs{\beta}(0)\coloneqq\bs{n}$. Thus, $\dd{}{\epsilon}\left(\bs{\alpha}(\epsilon)\cdot\bs{\alpha}(\epsilon)\right)\vert_{\epsilon=0}=2\bs{\alpha}^\prime(0)\cdot\bs{\nu}=0$ and $\dd{}{\epsilon}\left(\bs{\beta}(\epsilon)\cdot\bs{\beta}(\epsilon)\right)\vert_{\epsilon=0}=2\bs{\beta}^\prime(0)\cdot\bs{n}=0$, or $\bs{\alpha}^\prime(0)\perp\bs{\nu}$ and $\bs{\beta}^\prime(0)\perp\bs{n}$. Furthermore, $\bs{\alpha}^\prime(0)$ and $\bs{\beta}^\prime(0)$ can be specified as any vectors perpendicular to both $\bs{n}$ and $\bs{\nu}$. We then characterize the partial intrinsic derivatives of the boundary-edge microtraction
\begin{equation}
\dd{}{\epsilon}\tauds(\bs{x},t;\bs{\alpha}(\epsilon),\bs{\beta}(\epsilon))\Big\vert_{\epsilon=0}\eqqcolon({\tauds}_{\bs{\nu}}(\bs{x},t;\bs{\nu},\bs{n}))\cdot\bs{\alpha}^\prime(0)+({\tauds}_{\bs{n}}(\bs{x},t;\bs{\nu},\bs{n}))\cdot\bs{\beta}^\prime(0).
\end{equation}
Analyzing each component, we can state that
\begin{subequations}
\begin{align}
({\tauds}_{\bs{\nu}}(\bs{x},t;\bs{\nu},\bs{n}))\cdot\bs{\alpha}^\prime(0)&=({\tauds}_{\bs{\nu}}(\bs{x},t;\bs{\nu},\bs{n}))\cdot(\id-\bs{n}\otimes\bs{n}-\bs{\nu}\otimes\bs{\nu})\bs{\alpha}^\prime(0),\\
({\tauds}_{\bs{n}}(\bs{x},t;\bs{\nu},\bs{n}))\cdot\bs{\beta}^\prime(0)&=({\tauds}_{\bs{n}}(\bs{x},t;\bs{\nu},\bs{n}))\cdot(\id-\bs{n}\otimes\bs{n}-\bs{\nu}\otimes\bs{\nu})\bs{\beta}^\prime(0),
\end{align}
\end{subequations}
we then conclude that ${\tauds}_{\bs{\nu}}$ and ${\tauds}_{\bs{n}}$ align with $\bs{t}$, that is,
\begin{subequations}
\begin{align}
{\tauds}_{\bs{\nu}}(\bs{x},t;\bs{\nu},\bs{n})&=(\id-\bs{n}\otimes\bs{n}-\bs{\nu}\otimes\bs{\nu}){\tauds}_{\bs{\nu}}(\bs{x},t;\bs{\nu},\bs{n}),\\
{\tauds}_{\bs{n}}(\bs{x},t;\bs{\nu},\bs{n})&=(\id-\bs{n}\otimes\bs{n}-\bs{\nu}\otimes\bs{\nu}){\tauds}_{\bs{n}}(\bs{x},t;\bs{\nu},\bs{n}).
\end{align}
\end{subequations}
Thus,
\begin{equation}
{\tauds}_{\bs{\nu}}\cdot\bs{\nu}={\tauds}_{\bs{\nu}}\cdot\bs{n}={\tauds}_{\bs{n}}\cdot\bs{\nu}={\tauds}_{\bs{n}}\cdot\bs{n}=0.
\end{equation}
\qed
\end{rmk}

With Remark \ref{rmk:intrinsic.derivatives.edge} and the identities in \eqref{eq:identities}, the first variation of $\cl{F}$ reads
\begin{align}
\delta\cl{F}(\srf)[\bs{u}]=&\int\limits_{\dsrf}\left[\left({\tauds}_{\bs{x}}\dd{\bs{y}_\epsilon}{\epsilon}+{\tauds}_{\bs{\nu}}\dd{\bs{\nu}_\epsilon}{\epsilon}+{\tauds}_{\bs{n}}\dd{\bs{n}_\epsilon}{\epsilon}\right)|\bs{F}_\epsilon\bs{t}|+\tauds\dd{|\bs{F}_\epsilon\bs{t}|}{\epsilon}\right]_{\epsilon=0}\ds\nonumber\\
=&\int\limits_{\dsrf}\left(\tauds\pards{\bs{u}}{s}\cdot\bs{t}+{\tauds}_{\bs{x}}\cdot\bs{u}-({\tauds}_{\bs{\nu}}\cdot\bs{t})\pards{\bs{u}}{s}\cdot\bs{\nu}-{\tauds}_{\bs{n}}\cdot(\Grads(\bs{u}\cdot\bs{n})+\bs{L}\bs{P}\bs{u})\right)\ds\nonumber\\
=&\int\limits_{\dsrf}\left[\left({\tauds}_{\bs{x}}+\pards{}{s}\left(({\tauds}_{\bs{\nu}}\cdot\bs{t})\bs{\nu}+({\tauds}_{\bs{n}}\cdot\bs{t})\bs{n}-\tauds\bs{t}\right)\right)\cdot\bs{u}-{\tauds}_{\bs{n}}\cdot\bs{L}\bs{P}\bs{u}\right]\ds\nonumber\\
&-\int\limits_{\dsrf}\pards{}{s}\left[\left(({\tauds}_{\bs{\nu}}\cdot\bs{t})\bs{\nu}+({\tauds}_{\bs{n}}\cdot\bs{t})\bs{n}-\tauds\bs{t}\right)\cdot\bs{u}\right]\ds,
\end{align}
where the conditions ${\tauds}_{\bs{n}}=(\id-\bs{n}\otimes\bs{n}-\bs{\nu}\otimes\bs{\nu}){\tauds}_{\bs{n}}$ and ${\tauds}_{\bs{\nu}}=(\id-\bs{n}\otimes\bs{n}-\bs{\nu}\otimes\bs{\nu}){\tauds}_{\bs{\nu}}$ are used (cf. remark \ref{rmk:intrinsic.derivatives.edge}). The integral $\int_{\dsrf}\pards{}{s}(\cdot)\ds$ represents a jump condition at $\bs{x}\in\dsrf$ if the $\bs{t}$ is discontinuous at $\bs{x}$. We thus can explicitly write
\begin{equation}\label{eq:jump.edge.microtraction}
\llbracket({\tauds}_{\bs{\nu}}\cdot\bs{t})\bs{\nu}+({\tauds}_{\bs{n}}\cdot\bs{t})\bs{n}-\tauds\bs{t}\rrbracket=0,
\end{equation}
at $\bs{x}$, where $\llbracket\cdot\rrbracket=(\cdot)^+-(\cdot)^-$ implies a jump condition, `before-after' the discontinuty of $\bs{t}$ following the direction of $\bs{t}$. To unfold the implications of the jump \eqref{eq:jump.edge.microtraction}, two particular scenarios are analyzed as follow.

\begin{rmk}[\textbf{\textit{Scenarios of a nonsmooth edge $\dsrf$}}]\label{rmk:scenarios.nonsmooth.edge}
There are two particular scenarios to be considered when it comes to analyzing a nonsmooth edge $\dsrf$. Consider a unit tangent $\bs{t}$ discontinuous at some $\bs{x}$ of $\dsrf$. Then, in the first scenario, the unit normal $\bs{n}$ is continuous and the unit tangent-normal $\bs{\nu}$ is discontinuous, while alternatively in the second scenario, $\bs{n}$ is discontinuous but the unit tangent-normal $\bs{\nu}$ is continuous.

To illustrate both scenarios, one can imagine two particular cases. In the first scenario, imagine a square surface. The unit normal $\bs{n}$ is continuous everywhere, but at the corners the unit tangent $\bs{t}$ and the unit tangent-normal $\bs{\nu}$ are discontinuous. In the second scenario, imagine half of a cone. The unit tangent-normal $\bs{\nu}$ is continuous everywhere, but at its vertex the unit normal $\bs{n}$ and the unit tangent $\bs{t}$ are discontinuous.
\qed
\end{rmk}

Bearing in mind Remarks \ref{rmk:intrinsic.derivatives.edge} and \ref{rmk:scenarios.nonsmooth.edge}, we are led to speculate about the existence of a hypermicrostress tensor field as follows.

\begin{thm}[\textbf{\textit{Existence of a hypermicrostress tensor field}}]\label{th:existence.hypermicrostress}
The interaction between an edge $\dsrf$ of a smooth open oriented surface $\srf$ and the adjacent parts of $\bdy\setminus\srf$ invokes the existence of a linear transformation ${\bs{\mt{\Sigma}}(\bs{x},t)\in\mathrm{Lin}}$\footnote{`$\mathrm{Lin}$' is a vector space $\cl{V}\times\cl{V}$ where all elements $\bs{V}$ are linear transformations.}, denoted as the hypermicrostress tensor field in $\bdy$ for all $(\bs{x},t)$.
\end{thm}
\begin{proof}
Consider the two particular scenarios on nonsmooth edge $\dsrf$ described in Remark \ref{rmk:scenarios.nonsmooth.edge}:
\begin{enumerate}[label=(\roman*),font=\itshape,leftmargin=*]
\item First Scenario: $\bs{t}$ and $\bs{\nu}$ are discontinuous at some $\bs{x}$ of $\dsrf$ while $\bs{n}$ is continuous;
\item Second Scenario: $\bs{t}$ and $\bs{n}$ are discontinuous at some $\bs{x}$ of $\dsrf$ while $\bs{\nu}$ is continuous.
\end{enumerate}
The inner product of the jump \eqref{eq:jump.edge.microtraction} with $\bs{t}^+$ yields, in the first scenario,
\begin{equation}
\tauds(\bs{x},t;\bs{\nu}^+,\bs{n}) = -\left[({\tauds}_{\bs{\nu}}(\bs{x},t;\bs{\nu}^-,\bs{n})\cdot\bs{t}^-)\bs{\nu}^-\cdot\bs{t}^+-\tauds(\bs{x},t;\bs{\nu}^-,\bs{n})\bs{t}^-\cdot\bs{t}^+\right],
\end{equation}
and considering that $\bs{t}^+\cdot\bs{\nu}^-=\bs{t}^-\cdot\bs{\nu}^+$ and $\bs{t}^+\cdot\bs{t}^-=\bs{\nu}^-\cdot\bs{\nu}^+$, we arrive at
\begin{equation}
\tauds(\bs{x},t;\bs{\nu}^+,\bs{n}) = -\left[({\tauds}_{\bs{\nu}}(\bs{x},t;\bs{\nu}^-,\bs{n})\cdot\bs{t}^-)\bs{t}^--\tauds(\bs{x},t;\bs{\nu}^-,\bs{n})\bs{\nu}^-\right]\cdot\bs{\nu}^+,
\end{equation}
whereas, in the second scenario,
\begin{equation}
\tauds(\bs{x},t;\bs{\nu},\bs{n}^+) = -\left[({\tauds}_{\bs{n}}(\bs{x},t;\bs{\nu},\bs{n}^-)\cdot\bs{t}^-)\bs{n}^-\cdot\bs{t}^+-\tauds(\bs{x},t;\bs{\nu},\bs{n}^-)\bs{t}^-\cdot\bs{t}^+\right],
\end{equation}
and considering that $\bs{t}^+\cdot\bs{n}^-=\bs{t}^-\cdot\bs{n}^+$ and $\bs{t}^+\cdot\bs{t}^-=\bs{n}^-\cdot\bs{n}^+$, we arrive at
\begin{equation}
\tauds(\bs{x},t;\bs{\nu},\bs{n}^+) = -\left[({\tauds}_{\bs{n}}(\bs{x},t;\bs{\nu},\bs{n}^-)\cdot\bs{t}^-)\bs{t}^--\tauds(\bs{x},t;\bs{\nu},\bs{n}^-)\bs{n}^-\right]\cdot\bs{n}^+.
\end{equation}
In the first scenario, by keeping $\bs{t}^-$ fixed and consequently $\bs{\nu}^-$ fixed as well, the boundary-edge microtraction $\tauds$ is linear in $\bs{\nu}$, that is,
\begin{equation}\label{eq:linear.nu}
\tauds(\bs{x},t;\bs{\nu},\bs{n})\coloneqq\bs{a}\cdot\bs{\nu},\qquad\bs{a}\coloneqq\bs{a}(\bs{x},t;\bs{n}),\qquad\text{and}\qquad\bs{a}\cdot\bs{n}=0,\qquad\forall\,\bs{n},\bs{\nu}\in\mathrm{Unit},\quad\bs{n}\cdot\bs{\nu}=0,
\end{equation}
alternatively, in the second scenario, by keeping $\bs{t}^-$ fixed and consequently $\bs{n}^-$ fixed as well, the boundary-edge microtraction $\tauds$ is linear on $\bs{n}$, that is,
\begin{equation}\label{eq:linear.n}
\tauds(\bs{x},t;\bs{\nu},\bs{n})\coloneqq\bs{b}\cdot\bs{n},\qquad\bs{b}\coloneqq\bs{b}(\bs{x},t;\bs{\nu}),\qquad\text{and}\qquad\bs{b}\cdot\bs{\nu}=0,\qquad\forall\,\bs{n},\bs{\nu}\in\mathrm{Unit},\quad\bs{n}\cdot\bs{\nu}=0.
\end{equation}
To encompass conditions \eqref{eq:linear.nu} and \eqref{eq:linear.n} into a single one, we let $\{\bs{e}_1,\bs{e}_2,\bs{e}_3\coloneqq\bs{n}\}$ be a orthonormal basis and drop the dependency on $\bs{x}$ and $t$. Combining \eqref{eq:linear.nu}$_1$ and \eqref{eq:linear.n}$_1$, we have that
\begin{align}\label{eq:boundary.edge.microtraction.components}
(\bs{a}(\bs{n})\cdot\bs{e}_\alpha)\bs{e}_\alpha&=(\bs{b}(\bs{e}_\alpha)\cdot\bs{n})\bs{e}_\alpha,\nonumber\\
&=(\bs{e}_\alpha\otimes\bs{b}(\bs{e}_\alpha))\bs{n},
\end{align}
where the index $\alpha$ goes from $1$ to $2$, leaving out the unit normal $\bs{e}_3=\bs{n}$ from the set $\bs{e}_\alpha$.
Now, noting that \eqref{eq:boundary.edge.microtraction.components} is $\bs{a}(\bs{n})$, the boundary-edge microtraction $\tauds(\bs{x},t;\bs{\nu},\bs{n})=\bs{a}(\bs{n})\bs{\nu}$ can be specified as
\begin{equation}
\tauds(\bs{x},t;\bs{\nu},\bs{n})=\bs{\nu}\cdot(\bs{e}_\alpha\otimes\bs{b}(\bs{e}_\alpha))\bs{n}.
\end{equation}
Next, we express $\bs{b}(\bs{e}_\alpha)$ in a fixed orthonormal basis $\bs{e}^\prime_i$, such that $\bs{b}(\bs{e}_\alpha)=b^\prime_i(\bs{e}_\alpha)\bs{e}^\prime_i$. By using this generic but fixed orthonormal basis, the the boundary-edge traction $\tauds(\bs{x},t;\bs{\nu},\bs{n})$ assumes the forms
\begin{align}
\tauds(\bs{x},t;\bs{\nu},\bs{n})=\bs{\nu}\cdot(\underbrace{b^\prime_i(\bs{e}_\alpha)\bs{e}_\alpha\otimes\bs{e}^\prime_i}_{\bs{\mt{\Sigma}}(\bs{x},t)})\bs{n}.
\end{align}
Therefore, based on the jump condition \eqref{eq:jump.edge.microtraction} there exists a linear transformation ${\bs{\mt{\Sigma}}(\bs{x},t)\in\mathrm{Lin}}$, denoted as the hypermicrostress tensor field in $\bdy$ for all $(\bs{x},t)$, such that
\begin{equation}\label{eq:boundary.edge.microtraction}
\tauds(\bs{x},t;\bs{\nu},\bs{n})\coloneqq\bs{\nu}\cdot\bs{\mt{\Sigma}}(\bs{x},t)\bs{n},
\end{equation}
\end{proof}

Next, for the surface microtraction, we characterize a jump condition throughout the surface based on a surface balance of microforces and Theorem \ref{th:existence.hypermicrostress} as follows.
\begin{prop}[\textbf{\textit{Surface microtraction: jump condition throughout the surface}}]\label{pp:jump.surface.microtraction.smooth.open.S}
Consider a smooth open surface $\srf$ oriented by an unit normal $\bs{n}$ with boundary-edge $\dsrf$. Let $\xis\coloneqq\xis(\bs{x},t;\bs{n},\bs{L})$ and $\xis^\ast\coloneqq(\bs{x},t;\bs{n},\bs{L})$ be the surface microtractions defined on opposite sides of $\srf$ and $\tauds\coloneqq\tauds(\bs{x},t;\bs{\nu},\bs{n})$ the boundary-edge microtraction. In balancing these microforces on $\srf$ while accounting for \eqref{eq:boundary.edge.microtraction} from Theorem \ref{th:existence.hypermicrostress}, the following jump condition is obtained.
\begin{equation}\label{eq:surface.microtraction.opposite}
-\xis^\ast=\xis+\Divs(\bs{P}\mskip-2.5mu\bs{\mt{\Sigma}}\bs{n}),
\end{equation}
\end{prop}
\begin{proof}
We postulate the surface balance of microforces on a smooth surface $\srf$ as follows
\begin{equation}\label{eq:surface.microforce.balance.smooth.open.S}
\int\limits_{\srf}(\xis+\xis^\ast)\da+\int\limits_{\dsrf}\tauds\ds=0,\quad\forall\,\srf\subset\bdy\quad\text{and}\quad t.
\end{equation}
Next, consider the surface divergence theorem on a smooth open surface $\srf$,
\begin{equation}\label{eq:smooth.divs.theo.open.S}
\int\limits_{\srf}\Divs(\bs{P}\bs{g})\da=\int\limits_{\dsrf}\bs{g}\cdot\bs{\nu}\ds,
\end{equation}
for any vector field $\bs{g}$ on $\srf$. Thus, with the expression for the boundary-edge microtraction \eqref{eq:boundary.edge.microtraction} in the surface balance of microforces \eqref{eq:surface.microforce.balance.smooth.open.S} and applying the surface divergence theorem for smooth open surfaces, we are led to
\begin{equation}\label{eq:partwise.jump.surface.microtraction}
\int\limits_{\srf}(\xis+\xis^\ast+\Divs(\bs{P}\mskip-2.5mu\bs{\mt{\Sigma}}\bs{n}))\da=0.
\end{equation}
While by localizing it, we arrive at the statement of this proposition, where $\Divs(\bs{P}\mskip-2.5mu\bs{\mt{\Sigma}}\bs{n})$ represents a jump condition throughout the surface $\srf$ from one side to the other.
\end{proof}

Now, recalling the variational setting of the proof of the Cauchy theorem \cite{Fos89}, we establish that there exists a microstress-like field $\bs{\zeta}$ such that
\begin{equation}
-\xis^\ast(\bs{x},t;\bs{n},\bs{L})=\bs{\zeta}(\bs{x},t)\cdot\bs{n},
\end{equation}
while the surface microtraction \eqref{eq:surface.microtraction.opposite} becomes
\begin{equation}\label{eq:surface.microtraction.zeta}
\xis(\bs{x},t;\bs{n},\bs{L})=\bs{\zeta}(\bs{x},t)\cdot\bs{n}-\Divs(\bs{P}\mskip-2.5mu\bs{\mt{\Sigma}}\bs{n}).
\end{equation}
Note that, in the absence of higher-order effects, such as the hypermicrostress $\bs{\mt{\Sigma}}$, we recover the microtraction presented by Fried \& Gurtin \cite{Fri93}, that is, $\xis(\bs{x},t;\bs{n})=\bs{\zeta}(\bs{x},t)\cdot\bs{n}$.

\subsubsection{Boundary-edge microtraction characterization and its implications: part 2}

In analyzing the first integral in \eqref{eq:functional.H.epsilon}, for the point we wish to make, it is enough to analyze ${\xis}$ instead of ${\xis}+{\xis^\ast}$. So, let
\begin{equation}
\cl{G}(\srf_\epsilon)\coloneqq\int\limits_{\srf_\epsilon}{\xis}_\epsilon\da_\epsilon=\int\limits_{\srf}\xis(\bs{y}_\epsilon,t;\bs{n}_\epsilon,\bs{L}_\epsilon)J|\bs{F}^{-\trans}_\epsilon\bs{n}|\da,
\end{equation}
be a functional on $\srf_\epsilon$. Additionally, let ${\xis}_{\bs{x}}$, ${\xis}_{\bs{n}}$, and ${\xis}_{\bs{L}}$ be the intrinsic partial derivatives of $\xis$ with respect to $\bs{x}$, $\bs{n}$, and $\bs{L}$, respectively. The partial derivative ${\xis}_{\bs{x}}$ need not any additional characterization. The remainder derivatives, ${\xis}_{\bs{n}}$ and ${\xis}_{\bs{L}}$ need further understanding as follows.

\begin{rmk}[\textbf{\textit{Characterization of the partial intrinsic derivatives of the surface microtraction}}]\label{rmk:intrinsic.derivatives.surf}
Let $\bs{\beta}(\epsilon)$ and $\bs{B}(\epsilon)$ be smooth parametric orthonormal-vector- and symmetric-tensor-valued functions of $\epsilon\in[0,1]$, that is, $\bs{\beta}(\epsilon)\colon[0,1]\mapsto\mathrm{Unit}$, $\bs{B}(\epsilon)\colon[0,1]\mapsto\mathrm{Sym}$\footnote{`$\mathrm{Sym}$' is a vector space $\cl{V}\times\cl{V}$ where all elements $\bs{V}$ are symmetric tensors, that is, invariant under a permutation of its vector arguments.}, and $\bs{B}(\epsilon)\bs{\beta}(\epsilon)=\bs{0}$ $\forall\,\epsilon\in[0,1]$, where $\bs{\beta}(0)\coloneqq\bs{n}$ and $\bs{B}(0)\coloneqq\bs{L}$. Thus, $\dd{}{\epsilon}\left(\bs{B}(\epsilon)\cdot\bs{\beta}(\epsilon)\right)\vert_{\epsilon=0}=\bs{B}^\prime(0)\bs{n}+\bs{L}\bs{\beta}^\prime(0)=\bs{0}$. We then characterize the partial intrinsic derivatives of the surface microtraction as
\begin{equation}
\dd{}{\epsilon}\xis(\bs{x},t;\bs{\beta}(\epsilon),\bs{B}(\epsilon))\Big\vert_{\epsilon=0}\eqqcolon({\xis}_{\bs{n}}(\bs{x},t;\bs{n},\bs{L}))\cdot\bs{\beta}^\prime(0)+({\xis}_{\bs{L}}(\bs{x},t;\bs{n},\bs{L}))\colon\bs{B}^\prime(0).
\end{equation}
Analyzing each component, we conclude that
\begin{equation}
{\xis}_{\bs{n}}\cdot\bs{n}=0,\qquad{\xis}_{\bs{L}}\cdot\bs{n}=0,\qquad\text{and}\qquad{\xis}_{\bs{L}}\colon\bs{W}=0,\quad\forall\,\bs{W}\in\mathrm{Skw}.\footnote{`$\mathrm{Skw}$' is a vector space $\cl{V}\times\cl{V}$ where all elements $\bs{V}$ are skew-symmetric tensors, that is, under a permutation of its vector arguments the result is the opposite.}
\end{equation}
\qed
\end{rmk}

Thus, the first variation of $\cl{G}(\srf_\epsilon)$ is
\begin{align}
\delta\cl{G}(\srf)[\bs{u}]=&\int\limits_{\srf}\left[\left({\xis}_{\bs{x}}\dd{\bs{y}_\epsilon}{\epsilon}+{\xis}_{\bs{n}}\dd{\bs{n}_\epsilon}{\epsilon}+{\xis}_{\bs{L}}\dd{\bs{L}_\epsilon}{\epsilon}\right)J|\bs{F}^{-\trans}_\epsilon\bs{n}|+\xis\dd{J|\bs{F}^{-\trans}_\epsilon\bs{n}|}{\epsilon}\right]_{\epsilon=0}\da\nonumber\\
=&\int\limits_{\srf}\Big(\xis\mskip+2.5mu\Divs\bs{u}+{\xis}_{\bs{x}}\cdot\bs{u}-{\xis}_{\bs{n}}\cdot(\Grads\bs{u})^{\trans}\bs{n}\nonumber\\
&+{\xis}_{\bs{L}}\colon\left(\Grads\left((\Grads\bs{u})^{\trans}\bs{n}\right)-\bs{L}\mskip+2.5mu\Grads\bs{u}+\bs{L}(\Grads\bs{u})^{\trans}\bs{n}\otimes\bs{n}\right)\Big)\da.
\end{align}
Consider the surface divergence theorem
\begin{equation}\label{eq:divs.theo.openS}
\int\limits_{\srf}\Divs(\bs{P}\bs{g})\da=\int\limits_{\dsrf}\bs{g}\cdot\bs{\nu}\ds,
\end{equation}
for any vector field $\bs{g}$ on $\srf$, and the identity
\begin{equation}\label{eq:identity.div.surf}
\Divs\bs{g}=\Divs(\bs{P}\bs{g})+\Divs((\bs{g}\cdot\bs{n})\bs{n})=\Divs(\bs{P}\bs{g})-2K\bs{g}\cdot\bs{n},
\end{equation}
with $K\coloneqq-\fr{1}{2}\Divs\bs{n}$. Using \eqref{eq:divs.theo.openS} and \eqref{eq:identity.div.surf}, and the partial intrinsic derivatives from remark \ref{rmk:intrinsic.derivatives.surf}, we arrive at
\begin{align}
\delta\cl{G}(\srf)[\bs{u}]=&\int\limits_{\srf}\left({\xis}_{\bs{x}}-\Grads\xis-2\xis K\bs{n}+\Divs(\bs{n}\otimes{\xis}_{\bs{n}})+\Divs(\bs{n}\otimes\Divs{\xis}_{\bs{L}})+\Divs(({\xis}_{\bs{L}}\bs{L})^{\trans})\right)\cdot\bs{u}\da\nonumber\\
&+\int\limits_{\dsrf}\left[\left(\xis\bs{\nu}-({\xis}_{\bs{n}}\cdot\bs{\nu})\bs{n}-(\Divs{\xis}_{\bs{L}}\cdot\bs{\nu})\bs{n}-({\xis}_{\bs{L}}\bs{L})^{\trans}\bs{\nu}\right)\cdot\bs{u}+({\xis}_{\bs{L}}\bs{\nu})\cdot(\Grads\bs{u})^{\trans}\bs{n}\right]\ds.
\end{align}
Although this integral does have a contribution on $\dsrf$, it does not characterize a jump condition. Thus, the surface microtraction does not affect the proof of \eqref{eq:boundary.edge.microtraction}.

\subsubsection{Surface balance of microtorques}

With the boundary-edge microtraction expression \eqref{eq:boundary.edge.microtraction} and the surface microtraction jump \eqref{eq:surface.microtraction.opposite}, we determine the surface-couple microtraction jump across a smooth open oriented surface $\srf$ as follows.
\begin{prop}[\textbf{\textit{Surface-couple microtraction: jump condition throughout the surface}}]\label{pp:balance.microtorques.smooth.open.S}
Consider a smooth open surface $\srf$ with boundary $\dsrf$. Let $\bsvarpis\coloneqq\bsvarpis(\bs{x},t;\bs{n})$ and $\bsvarpis^\ast\coloneqq\bsvarpis(\bs{x},t;\bs{n})$ be the surface-couple microtractions defined on opposite sides of $\srf$. In balancing these microtorques with the microtorques provoked by all the microtractions on $\srf$ while accounting for \eqref{eq:boundary.edge.microtraction} from Theorem \ref{th:existence.hypermicrostress} and \eqref{eq:surface.microtraction.opposite} from Proposition \ref{pp:jump.surface.microtraction.smooth.open.S}, the following jump condition is obtained,
\begin{equation}\label{eq:surface.couple.microtraction.opposite}
-\bsvarpis^\ast=\bsvarpis+\bs{P}\mskip-2.5mu\bs{\mt{\Sigma}}\bs{n},
\end{equation}
\end{prop}
\begin{proof}
We postulate the surface balance of microtorques on $\srf$ as follows
\begin{equation}\label{eq:surface.microtorque.balance.smooth.open.S}
\int\limits_{\srf}(\bsvarpis+\bsvarpis^\ast)\da+\int\limits_{\srf}(\xis+\xis^\ast)\bs{r}\da+\int\limits_{\dsrf}\tauds\bs{r}\ds=0,\quad\forall\,\srf\subset\bdy\quad\text{and}\quad t,
\end{equation}
being $\bs{r}\coloneqq\bs{x}-\bs{o}$ a vector, where $\bs{o}$ is a fixed reference point.

Using the boundary-edge microtraction \eqref{eq:boundary.edge.microtraction} and the jump condition for surface microtraction \eqref{eq:surface.microtraction.opposite}, the surface balance of microtorques \eqref{eq:surface.microtorque.balance.smooth.open.S} becomes
\begin{equation}
\int\limits_{\srf}(\bsvarpis+\bsvarpis^\ast)\da-\int\limits_{\srf}\bs{r}\mskip+2.5mu\Divs(\bs{P}\mskip-2.5mu\bs{\mt{\Sigma}}\bs{n})\da+\int\limits_{\dsrf}\bs{r}(\bs{\nu}\cdot\bs{\mt{\Sigma}}\bs{n})\ds=0.
\end{equation}
Next, with the surface divergence theorem for smooth open surfaces \eqref{eq:smooth.divs.theo.open.S} and the identity
\begin{equation}
\Divs(\bs{r}\otimes\bs{P}\mskip-2.5mu\bs{\mt{\Sigma}}\bs{n})=\bs{P}\mskip-2.5mu\bs{\mt{\Sigma}}\bs{n}+\bs{r}\mskip+2.5mu\Divs(\bs{P}\mskip-2.5mu\bs{\mt{\Sigma}}\bs{n}),
\end{equation}
the surface balance of microtorques \eqref{eq:surface.microtorque.balance.smooth.open.S} specializes further to
\begin{equation}\label{eq:partwise.jump.surface.couple.microtraction}
\int\limits_{\srf}(\bsvarpis+\bsvarpis^\ast+\bs{P}\mskip-2.5mu\bs{\mt{\Sigma}}\bs{n})\da=0,
\end{equation}
and localizing it, to arrive at the statement of this proposition, where $\bs{P}\mskip-2.5mu\bs{\mt{\Sigma}}\bs{n}$ represents a jump condition across the surface $\srf$.
\end{proof}

\subsection{Analysis on a nonsmooth open oriented surface $\srf$}

Applying Fosdick's procedure \cite{Fos16}, we proved that the edge microtraction invokes the existence of a hypermicrostress field. Moreover, the edge microtraction is linear on $\bs{\nu}$ and $\bs{n}$ through the hypermicrostress. Next, we study the surface balance of microforces and microtorques on a nonsmooth open oriented surface $\srf$ with boundary $\dsrf$ while considering the lack of smoothness of $\srf$ along $\edg$.

\subsubsection{Surface balance of microforces}

Having characterized the boundary-edge microtraction \eqref{eq:boundary.edge.microtraction} and the surface microtraction jump \eqref{eq:surface.microtraction.opposite}, we determine the internal-edge microtraction on a nonsmooth open oriented surface $\srf$ as follows.
\begin{prop}[\textbf{\textit{Internal-edge microtraction}}]\label{pp:internal.edge.microtraction.nonsmooth.open.S}
Consider a nonsmooth open surface $\srf$ with boundary $\dsrf$ and an internal-edge $\edg$. Let $\xis\coloneqq\xis(\bs{x},t;\bs{n},\bs{L})$ and $\xis^\ast\coloneqq(\bs{x},t;\bs{n},\bs{L})$ be the surface microtractions defined on opposite sides of $\srf$, $\tauds\coloneqq\tauds(\bs{x},t;\bs{\nu},\bs{n})$ and $\tauc\coloneqq\tauc(\bs{x},t;\bs{n}^+,\bs{n}^-)$ the boundary- and internal-edge microtractions, respectively. In balancing these microforces on $\srf$ while accounting for \eqref{eq:boundary.edge.microtraction} from Theorem \ref{th:existence.hypermicrostress} and \eqref{eq:surface.microtraction.opposite} from Proposition \ref{pp:jump.surface.microtraction.smooth.open.S}, the following representation for the internal-edge microtraction is obtained,
\begin{equation}\label{eq:internal.edge.microtraction}
\tauc=\surp{\bs{\nu}\cdot\bs{\mt{\Sigma}}\bs{n}},
\end{equation}
where $\surp{\bs{\nu}\cdot\bs{\mt{\Sigma}}\bs{n}}\coloneqq\bs{\nu}^+\cdot\bs{\mt{\Sigma}}\bs{n}^++\bs{\nu}^-\cdot\bs{\mt{\Sigma}}\bs{n}^-$.
\end{prop}
\begin{proof}
We postulate the surface balance of microforces on $\srf$ as follows
\begin{equation}\label{eq:surface.microforce.balance.nonsmooth.open.S}
\int\limits_{\srf}(\xis+\xis^\ast)\da+\int\limits_{\dsrf}\tauds\ds+\int\limits_{\edg}\tauc\ds=0,\quad\forall\,\srf\subset\bdy\quad\text{and}\quad t.
\end{equation}

Owing to the lack of smoothness at an internal-edge $\edg$, the surface divergence theorem over a closed surface exhibits a \emph{surplus}, that is,
\begin{equation}\label{eq:nonsmooth.divs.theo.closed.S}
\int\limits_{\srf}\Divs(\bs{P}\bs{g})\da=\int\limits_{\edg}\surp{\bs{g}\cdot\bs{\nu}}\ds,
\end{equation}
where $\surp{\bs{g}\cdot\bs{\nu}}\coloneqq\bs{g}^+\cdot\bs{\nu}^++\bs{g}^-\cdot\bs{\nu}^-$, for any smooth vector field $\bs{g}$ on $\srf$ with limiting values $\bs{g}^+$ and $\bs{g}^-$ on $\edg$. For open surfaces, the surface divergence theorem \eqref{eq:nonsmooth.divs.theo.closed.S} reads
\begin{equation}\label{eq:nonsmooth.divs.theo.open.S}
\int\limits_{\srf}\Divs(\bs{P}\bs{g})\da=\int\limits_{\dsrf}\bs{g}\cdot\bs{\nu}\ds+\int\limits_{\edg}\surp{\bs{g}\cdot\bs{\nu}}\ds,
\end{equation}
for any vector field $\bs{g}$ on $\srf$.

By the surface divergence theorem \eqref{eq:nonsmooth.divs.theo.open.S} with the boundary-edge microtraction \eqref{eq:boundary.edge.microtraction} and the jump condition for the surface microtraction \eqref{eq:surface.microtraction.opposite} between opposite sides of $\srf$, the surface balance of microforces \eqref{eq:surface.microforce.balance.nonsmooth.open.S} specializes to
\begin{equation}
\int\limits_{\edg}\left(\tauc-\surp{\bs{\nu}\cdot\bs{\mt{\Sigma}}\bs{n}}\right)\ds=0,
\end{equation}
and localizing it, we arrive at the statement of this proposition.
\end{proof}

\subsection{Analysis on an arbitrary part $\prt$}

On an arbitrary part $\prt$, we postulate the partwise balance of microforces as follows
\begin{equation}\label{eq:linear.momentum}
\int\limits_{\prt}(\pi+\gamma)\dv+\int\limits_{\dprt}\xis\da+\int\limits_{\edg}\tauc\ds=0,\quad\forall\,\prt\subseteq\bdy\quad\text{and}\quad t,
\end{equation}
where $\pi$ and $\gamma$ are the internal and external microforces densities. We also postulate the partwise balance of microtorques on an arbitrary part $\prt$, which reads
\begin{equation}\label{eq:angular.momentum}
\int\limits_{\prt}\left((\pi+\gamma)\bs{r}-\bs{\xi}\right)\dv+\int\limits_{\dprt}(\xis\bs{r}+\bsvarpis)\da+\int\limits_{\edg}\tauc\bs{r}\ds=\bs{0},\quad\forall\,\prt\subseteq\bdy\quad\text{and}\quad t,
\end{equation}
where $\bs{\xi}$ is a microstress and $\bs{r}=\bs{x}-\bs{o}$ a vector for a fixed point $\bs{o}$.

\subsubsection{Pointwise microforce balance}

Substituting the internal-edge microtraction \eqref{eq:internal.edge.microtraction} and the surface microtration representation \eqref{eq:surface.microtraction.zeta} into the partwise balance of microforces \eqref{eq:linear.momentum}, and applying the surface divergence theorem for nonsmooth closed surfaces \eqref{eq:nonsmooth.divs.theo.closed.S}, we obtain
\begin{equation}
\int\limits_{\prt}(\pi+\gamma)\dv+\int\limits_{\dprt}\bs{\zeta}\cdot\bs{n}\da=0.
\end{equation}
Using the volume divergence theorem followed by localization, we obtain the following pointwise balance of microforces
\begin{equation}\label{eq:microforce.balance}
\Div\bs{\zeta}+\pi+\gamma=0.
\end{equation}

\subsubsection{Pointwise microtorque balance}

Substituting the internal-edge microtraction \eqref{eq:internal.edge.microtraction} and the surface microtration representation \eqref{eq:surface.microtraction.zeta} into the partwise balance of microtorques \eqref{eq:angular.momentum}, we obtain
\begin{equation}
\int\limits_{\prt}\left(\bs{r}(\pi+\gamma)-\bs{\xi}\right)\dv+\int\limits_{\dprt}\left(\bs{r}(\bs{\zeta}\cdot\bs{n}-\Divs(\bs{P}\mskip-2.5mu\bs{\mt{\Sigma}}\bs{n}))+\bsvarpis\right)\da+\int\limits_{\edg}\bs{r}\surp{\bs{\nu}\cdot\bs{\mt{\Sigma}}\bs{n}}\ds=\bs{0},
\end{equation}
while considering the identity,
\begin{equation}\label{eq:id.varpis.ast}
\Divs(\bs{r}\otimes\bs{P}\mskip-2.5mu\bs{\mt{\Sigma}}\bs{n})=\bs{\mt{\Sigma}}\bs{n}-(\bs{n}\otimes\bs{n})\bs{\mt{\Sigma}}\bs{n}+\bs{r}\mskip+2.5mu\Divs(\bs{P}\mskip-2.5mu\bs{\mt{\Sigma}}\bs{n}),
\end{equation}
and applying the surface divergence theorem for nonsmooth closed surfaces \eqref{eq:nonsmooth.divs.theo.closed.S}, we arrive at
\begin{equation}
\int\limits_{\prt}\left(\bs{r}(\pi+\gamma)-\bs{\xi}\right)\dv+\int\limits_{\dprt}\left((\bs{r}\otimes\bs{\zeta}+\bs{\mt{\Sigma}})\bs{n}+\bsvarpis-(\bs{n}\otimes\bs{n})\bs{\mt{\Sigma}}\bs{n}\right)\da=\bs{0}.
\end{equation}
Now, applying the volume divergence theorem, we obtain the following representation
\begin{equation}
\int\limits_{\prt}\bs{r}(\Div\bs{\zeta}+\pi+\gamma)\dv+\int\limits_{\prt}\left(\bs{\zeta}+\Div\bs{\mt{\Sigma}}-\bs{\xi}\right)\dv+\int\limits_{\dprt}\left(\bsvarpis-(\bs{n}\otimes\bs{n})\bs{\mt{\Sigma}}\bs{n}\right)\ds=\bs{0},
\end{equation}
where the first integral is zero by the pointwise balance of microforces \eqref{eq:microforce.balance}. Thus, the partwise balance of microtorques yields
\begin{equation}
\int\limits_{\prt}\left(\bs{\zeta}+\Div\bs{\mt{\Sigma}}-\bs{\xi}\right)\dv+\int\limits_{\dprt}\left(\bsvarpis-(\bs{n}\otimes\bs{n})\bs{\mt{\Sigma}}\bs{n}\right)\ds=\bs{0}.
\end{equation}
To annihilate the second integral, the surface-couple microtraction is assumed to have the form
\begin{equation}\label{eq:surface.couple.microtraction}
\bsvarpis=(\bs{n}\otimes\bs{n})\bs{\mt{\Sigma}}\bs{n},
\end{equation}
yielding the partwise balance of microtorques
\begin{equation}
\int\limits_{\prt}\left(\bs{\zeta}+\Div\bs{\mt{\Sigma}}-\bs{\xi}\right)\dv=\bs{0}.
\end{equation}
Localizing this expression, we arrive at the pointwise balance of microtorques
\begin{equation}\label{eq:microtorque.balance}
\bs{\zeta}=\bs{\xi}-\Div\bs{\mt{\Sigma}}.
\end{equation}
We can now rewrite the surface microtraction \eqref{eq:surface.microtraction.zeta} as
\begin{equation}\label{eq:surface.microtraction}
\xis=(\bs{\xi}-\Div\bs{\mt{\Sigma}})\cdot\bs{n}-\Divs(\bs{P}\mskip-2.5mu\bs{\mt{\Sigma}}\bs{n}).
\end{equation}
Furthermore, accounting for the pointwise microtorque balance \eqref{eq:microtorque.balance} into the pointwise balance of microforces \eqref{eq:microforce.balance}, we arrive at the field equation of the \emph{phase-field gradient theory}, which reads
\begin{equation}\label{eq:field.equation}
\Div(\bs{\xi}-\Div\bs{\mt{\Sigma}})+\pi+\gamma=0.
\end{equation}
Espath et al.\ \cite{Esp17,Esp20} also derived this equation. For convenience, we define the hypermicrotraction and its relation with the surface-couple microtraction
\begin{equation}\label{eq:hypermicrotraction}
\sigmas\coloneqq\bs{n}\cdot\bs{\mt{\Sigma}}\bs{n},\qquad\bsvarpis\coloneqq\sigmas\bs{n}.
\end{equation}

\begin{rmk}[\textbf{\textit{On the symmetry of the hypermicrostress}}]\label{rmk.sym.hypermicrostress}
There are some evidences that suggest that the hypermicrostress should be symmetric, and we list them as follows.
\begin{enumerate}[label=evidence (\roman*),font=\itshape,leftmargin=*]
\item Recalling that the internal-edge microtraction has the representation $\tauc=\surp{\bs{\nu}\cdot\bs{\mt{\Sigma}}\bs{n}}$ and since $\bs{n}^\pm$ and $\bs{\nu}^\pm$ live in the same plane, $(\bs{n}^+\otimes\bs{\nu}^++\bs{n}^-\otimes\bs{\nu}^-)\in\mathrm{Sym}$, only the symmetric part of the hypermicrostress contributes to the internal-edge microtraction, that is,
\begin{equation}
\surp{\bs{\nu}\cdot\bs{\mt{\Sigma}}\bs{n}}=\surp{\bs{\nu}\cdot\,\sym(\bs{\mt{\Sigma}})\bs{n}};
\end{equation}
\item Recalling that the surface-couple microtraction has the representation $\bsvarpis=(\bs{n}\otimes\bs{n})\bs{\mt{\Sigma}}\bs{n}$, only the symmetric part of the hypermicrostress contributes to it, that is,
\begin{equation}
(\bs{n}\otimes\bs{n})\bs{\mt{\Sigma}}\bs{n}=(\bs{n}\otimes\bs{n})\,\sym(\bs{\mt{\Sigma}})\bs{n};
\end{equation}
\item Given that $\bs{P}\mskip-2.5mu\,\skw(\bs{\mt{\Sigma}})\bs{n}=\skw(\bs{\mt{\Sigma}})\bs{n}$ and with the following identity
\begin{equation}
\Divs(\bs{A}^{\trans}\bs{n})=\bs{n}\cdot\Divs\bs{A}-\bs{A}\colon\bs{L},
\end{equation}
which holds for any tensor field $\bs{A}$ defined on $\srf$, then
\begin{equation}
\Divs(\skw(\bs{\mt{\Sigma}})\bs{n})=\bs{n}\cdot\Divs(\skw(\bs{\mt{\Sigma}})^{\trans})=-\bs{n}\cdot\Divs(\skw(\bs{\mt{\Sigma}}))
\end{equation}
and
\begin{equation}
\bs{n}\cdot(\bs{n}\otimes\bs{n}\colon(\Grad(\skw(\bs{\mt{\Sigma}})))=0,
\end{equation}
we obtain
\begin{equation}
-\bs{n}\cdot\Div(\skw(\bs{\mt{\Sigma}}))=\Divs(\bs{P}(\skw(\bs{\mt{\Sigma}}))\bs{n}).
\end{equation}
Thus, recalling that the surface microtraction has the representation $\xis=(\bs{\xi} - \Div\bs{\mt{\Sigma}})\cdot\bs{n} - \Divs(\bs{P}\mskip-2.5mu\bs{\mt{\Sigma}}\bs{n})$, only the symmetric part of the hypermicrostress contributes to it, that is,
\begin{equation}
(\bs{\xi} - \Div\bs{\mt{\Sigma}})\cdot\bs{n} - \Divs(\bs{P}\mskip-2.5mu\bs{\mt{\Sigma}}\bs{n}) = (\bs{\xi} - \Div(\sym(\bs{\mt{\Sigma}})))\cdot\bs{n} - \Divs(\bs{P}(\sym(\bs{\mt{\Sigma}}))\bs{n})
\end{equation}
\item Recalling that the field equation of the \emph{phase-field gradient theory} reads $\Div(\bs{\xi}-\Div\bs{\mt{\Sigma}})+\pi+\gamma=0$, only the symmetry part of the hypermicrostress contributes to it, since $\Div^2\bs{\mt{\Sigma}}=\Div^2(\sym(\bs{\mt{\Sigma}}))$.
\end{enumerate}
\qed
\end{rmk}

\subsubsection{Action-reaction principle}

An important consequence of the representations of the microtractions is that the surface microtraction and the surface-couple microtraction are local at any point $\bs{x}$ on $\srf$ and any time $t$ and that the internal-edge microtraction is local at any point $\bs{x}$ on $\edg$ and any time $t$. Moreover, we state that $\xis$ depends on $\srf$ through the unit normal $\bs{n}$ and the curvature tensor $\bs{L}$ of $\srf$ at $\bs{x}$, and $\bsvarpis$ depends on $\srf$ through $\bs{n}$ (cubically), whereas $\tauc$ depends on $\edg$ through the unit normals $\{\bs{n}^+,\bs{n}^-\}$ (or equivalently through the unit tangent-normals $\{\bs{\nu}^+,\bs{\nu}^-\}$).

Next, letting $-\srf$ ($-\edg$) denote the surface $\srf$ (internal-edge $\edg$) oriented by $-\bs{n}$ ($\{-\bs{n}^+,-\bs{n}^-\}$), with reference to \eqref{eq:curvature.tensor.L}, $-\srf$ has curvature tensor $-\bs{L}$, we see that
\begin{equation}\label{eq:action.reaction.micro}
\left\{\,
\begin{aligned}
\xis(\bs{x},t;-\bs{n},-\bs{L})&=-\xis(\bs{x},t;\bs{n},\bs{L}),\\[4pt]
\bsvarpis(\bs{x},t;-\bs{n})&=-\bsvarpis(\bs{x},t;\bs{n}),\\[4pt]
\tauc(\bs{x},t;-\bs{n}^+,-\bs{n}^-)&=-\tauc(\bs{x},t;\bs{n}^+,\bs{n}^-).
\end{aligned}
\right.
\end{equation}
The relations \eqref{eq:action.reaction.micro}$_{1,2}$ make explicit the action-reaction principle in terms of microtractions between two smooth surfaces endowed with opposite unit normals and opposite curvature tensors at a point. Conversely, the relation \eqref{eq:action.reaction.micro}$_3$ presents the action-reaction principle between two parts of the same oriented nonsmooth surface divided by an internal-edge, which in turn is defined by a pair of discontinuous unit normals at a point.

\section{Power balance}
\label{power.statement}

Once the theory is built upon balances of fundamental fields, the `principle of virtual powers' becomes a theorem. We thus state this theorem as follows.

\begin{thm}[\textbf{\textit{The virtual power theorem}}]\label{th:virtual.power.theorem}
Assuming that the external virtual power is expended by the following conjugate pairs
\begin{enumerate}[label=(\roman*),font=\itshape,leftmargin=*]
\item \emph{Surface microtraction power per unit area:} $\{\gamma,\chi\}$ on $\prt$;
\item \emph{Surface microtraction power per unit area:} $\{\xis,\chi\}$ on $\dprt$;
\item \emph{Surface-couple microtraction power per unit area:} $\{\bsvarpis,\bs{\omega}\}$ on $\dprt$, with $\bs{\omega}\coloneqq\Grad\chi$;
\item \emph{Internal-edge microtraction power per unit length:} $\{\tauc,\chi\}$ on $\edg\subset\dprt$,
\end{enumerate}
together with the assumption that the field equation \eqref{eq:field.equation} is satisfied, the virtual power balance
\begin{equation}
\int\limits_{\prt}(-\pi\chi+\bs{\xi}\cdot\Grad\chi+\bs{\mt{\Sigma}}\colon\Grad^2\chi)\dv=\int\limits_{\prt}\gamma\chi\dv+\int\limits_{\dprt}\Big(\xis\chi+\sigmas\pards{\chi}{n}\Big)\da+\int\limits_{\edg}\tauc\chi\ds,\qquad\forall\,\prt\subseteq\bdy\quad\text{and}\quad t,
\end{equation}
holds for any scalar smooth and admissible virtual field $\chi$.
\end{thm}
\begin{proof}
By using the hypermicrotraction representation \eqref{eq:hypermicrotraction}$_1$ of the surface-couple microtraction, the virtual power expended by pair $\{\bsvarpis,\bs{\omega}\}$ becomes
\begin{align}
   (\bsvarpis,\bs{\omega}) &= (\bsvarpis,\Grad\chi),\nonumber\\[4pt]
&= (\sigmas\bs{n},\Grad\chi),\nonumber\\[4pt]
&= (\sigmas,\Grad\chi\cdot\bs{n}),\nonumber\\[4pt]
&= (\sigmas,\partial{\chi}/\partial{n}).
\end{align}
Then, summing up the partwise power expenditures \emph{(i)-(iv)}, we arrive at the external virtual power
\begin{equation}\label{eq:external.virtual.power}
\cl{V}_{\mathrm{ext}}(\prt;\chi)\coloneqq\int\limits_{\prt}\gamma\chi\dv+\int\limits_{\dprt}\Big(\xis\chi+\sigmas\pards{\chi}{n}\Big)\da+\int\limits_{\edg}\tauc\chi\ds.
\end{equation}
Aided by the volume divergence and surface divergence theorems, and with the internal-edge microtraction \eqref{eq:internal.edge.microtraction}, surface-couple microtraction \eqref{eq:surface.couple.microtraction}, surface microtraction \eqref{eq:surface.microtraction}, and the field equation \eqref{eq:field.equation} of the \emph{phase-field gradient theory}, we are led from the external virtual power \eqref{eq:external.virtual.power} to the internal virtual power
\begin{equation}\label{eq:internal.virtual.power}
\cl{V}_{\mathrm{int}}(\prt;\chi)\coloneqq\int\limits_{\prt}(-\pi\chi+\bs{\xi}\cdot\Grad\chi+\bs{\mt{\Sigma}}\colon\Grad^2\chi)\dv.
\end{equation}

The virtual power balance, that is, the balance between the external and internal powers,
\begin{equation}\label{eq:virtual.power.balance}
\cl{V}_{\mathrm{ext}}(\prt;\chi)=\cl{V}_{\mathrm{int}}(\prt;\chi)
\end{equation}
holds for any arbitrary part $\prt$ and any choice of the virtual field $\chi$.
\end{proof}

Note that the actual power expenditures are
\begin{equation}\label{eq:external.power}
\cl{W}_{\mathrm{ext}}(\prt)\coloneqq\int\limits_{\prt}\gamma\dot{\varphi}\dv+\int\limits_{\dprt}\Big(\xis\dot{\varphi}+\sigmas\pards{\dot{\varphi}}{n}\Big)\da+\int\limits_{\edg}\tauc\dot{\varphi}\ds,
\end{equation}
and
\begin{equation}\label{eq:internal.power}
\cl{W}_{\mathrm{int}}(\prt)\coloneqq\int\limits_{\prt}(-\pi\dot{\varphi}+\bs{\xi}\cdot\Grad\dot{\varphi}+\bs{\mt{\Sigma}}\colon\Grad^2\dot{\varphi})\dv.
\end{equation}

\section{Thermodynamics}
\label{thermodynamics}

\subsection{First and second laws of thermodynamics}

The first two laws of thermodynamics for a continuum consist of balance of energy and an entropy imbalance that is frequently referred to as the Clausius--Duhem inequality. Following Truesdell \& Noll \cite[\S79]{Tru04}, these laws have the respective forms\footnote{The idea of using a virtual-power principle to generate an appropriate form of the external power expenditure in the energy balance was originated by Gurtin \cite[\S6]{Gur02}.}
\begin{equation}\label{eq:partwise.thermodynamical.laws}
\left\{\,
\begin{aligned}
\dot{\overline{\int\limits_{\prt}\varepsilon\dv}}&=\cl{W}_{\mathrm{ext}}(\prt)-\int\limits_{\dprt}\bs{q}\cdot\bs{n}\da+\int\limits_{\prt}r\dv,\\[4pt]
\dot{\overline{\int\limits_{\prt}\eta\dv}}&\ge-\int\limits_{\dprt}\dfrac{\bs{q}}{\vartheta}\cdot\bs{n}\da+\int\limits_{\prt}\dfrac{r}{\vartheta}\dv,
\end{aligned}
\right.
\end{equation}
where $\varepsilon$ and $\eta$ represent the internal-energy density and entropy density, $\bs{q}$ is the heat flux, $r$ is the heat supply, and $\vartheta>0$ is the absolute temperature.

Since, by \eqref{eq:virtual.power.balance}, $\cl{W}_{\mathrm{ext}}(\prt)=\cl{W}_{\mathrm{int}}(\prt)$, we may substitute the expression \eqref{eq:external.virtual.power} defining the power expended on $\prt$ by external agencies on the right-hand side of the energy balance \eqref{eq:partwise.thermodynamical.laws}$_1$ by the expression \eqref{eq:internal.virtual.power} defining the internal power of $\prt$. The result of localizing both of \eqref{eq:partwise.thermodynamical.laws} is
\begin{equation}\label{eq:pointwise.thermodynamical.laws}
\left\{\,
\begin{aligned}
\dot\varepsilon&=-\pi\dot\varphi+\bs{\xi}\cdot\Grad\dot\varphi+\bs{\mt{\Sigma}}\colon\Grad^2\dot\varphi-\Div\bs{q}+r,\\[4pt]
\dot\eta&\ge-\Div\dfrac{\bs{q}}{\vartheta}+\dfrac{r}{\vartheta}.
\end{aligned}
\right.
\end{equation}
Important to what follows is the free-energy density
\begin{equation}\label{eq:free.energy.def}
\psi\coloneqq\varepsilon-\vartheta\eta.
\end{equation}
Then, since \eqref{eq:pointwise.thermodynamical.laws}$_2$ may be written as
\begin{equation}
\dot\eta\ge-\dfrac{1}{\vartheta}\Div\bs{q}+\dfrac{1}{\vartheta^2}\bs{q}\cdot\Grad\vartheta+\dfrac{r}{\vartheta},
\end{equation}
if we multiply this equation by $\vartheta$ and subtract it from \eqref{eq:pointwise.thermodynamical.laws}$_1$, we arrive at the pointwise free-energy imbalance
\begin{equation}\label{eq:pointwise.free.energy.imbalance}
\dot\psi+\eta\dot{\vartheta}+\pi\dot\varphi-\bs{\xi}\cdot\Grad\dot\varphi-\bs{\mt{\Sigma}}\colon\Grad^2\dot\varphi+\dfrac{1}{\vartheta}\bs{q}\cdot\Grad\vartheta\le0.
\end{equation}

\subsection{Pointwise and partwise free-energy imbalances for isothermal processes}

Applications in which thermal changes are negligible are encompassed by the present framework when attention is restricted to isothermal processes, namely to processes in which
\begin{equation}
\vartheta=\vartheta_0\equiv\mathrm{constant}.
\end{equation}
For such processes, the expression \eqref{eq:free.energy.def} for the free-energy density specializes to
\begin{equation}
\psi=\varepsilon-\vartheta_0\eta
\end{equation}
and the pointwise free-energy imbalance \eqref{eq:pointwise.free.energy.imbalance} has the simple form
\begin{equation}\label{eq:pointwise.free.energy.imbalance.isothermal}
\dot\psi+\pi\dot\varphi-\bs{\xi}\cdot\Grad\dot\varphi-\bs{\mt{\Sigma}}\colon\Grad^2\dot\varphi\le0.
\end{equation}
Further, if we multiply \eqref{eq:pointwise.thermodynamical.laws}$_2$ by $\vartheta_0$ and subtract it from \eqref{eq:pointwise.thermodynamical.laws}$_1$ we arrive at a partwise free-energy imbalance
\begin{equation}\label{eq:partwise.imbalance}
\dot{\overline{\int\limits_{\prt}\psi\dv}}\le\cl{W}_{\mathrm{ext}}(\prt)
\end{equation}
requiring that the temporal increase in free energy of $\prt$ be less than or equal to the power expended by external agencies on $\prt$.

The imbalance \eqref{eq:partwise.imbalance} is often the implicit starting point of most purely mechanical theories. Of course, the pointwise imbalance \eqref{eq:pointwise.free.energy.imbalance.isothermal} may be derived as a direct consequence of \eqref{eq:partwise.imbalance}, the power balance \eqref{eq:virtual.power.balance}, and the expression \eqref{eq:internal.virtual.power} for the internal power without introducing the notion of temperature; in that sense, the imbalance stands on its own as a starting point for the development of purely mechanical theories.

\section{Nonconserved second-grade phase field equation}
\label{nonconserved.case}

We hereafter restrict attention to purely mechanical processes governed by the isothermal version \eqref{eq:pointwise.free.energy.imbalance.isothermal} of the pointwise free-energy imbalance. Guided by the presence of the power conjugate pairings $\pi\dot\varphi$, $\bs{\xi}\cdot\Grad\dot\varphi$, and $\bs{\mt{\Sigma}}\colon\Grad^2\dot\varphi$ in that inequality, we consider a class of constitutive equations that delivers the free-energy density $\psi$, internal microforce $\pi$, microstress $\bs{\xi}$, and hypermicrostress $\bs{\mt{\Sigma}}$ at each point $\bs{x}$ in $\bdy$ and each instant $t$ of time in terms of the values of the phase field $\varphi$, its first and second gradients $\Grad\varphi$ and $\Grad^2\varphi$, and its time rate $\dot\varphi$ at that point and time.

We do not prescribe a constitutive equation for the external microforce $\gamma$ but instead allow it to be chosen in any way that is needed to ensure the satisfaction of the field equation \eqref{eq:field.equation}. Arguments introduced by Coleman \& Noll \cite{Col63} can then be adapted to show that for the dissipation inequality \eqref{eq:pointwise.free.energy.imbalance.isothermal} to be satisfied in all processes it is necessary and sufficient to require that:
\begin{itemize}
\item The free-energy density $\psi$ is given by a constitutive response function $\hat{\psi}$ that is independent of $\dot\varphi$:
\begin{equation}\label{eq:free.energy.functional}
\psi=\hat{\psi}(\varphi,\Grad\varphi,\Grad^2\varphi).
\end{equation}
\item The microstress $\bs{\xi}$ and hypermicrostress $\bs{\mt{\Sigma}}$ are given by constitutive response functions $\hat{\bs{\xi}}$ and $\hat{\bs{\mt{\Sigma}}}$ that derive from the response function $\hat{\psi}$:
\begin{equation}\label{eq:xi.Sigma.constitutive.relation}
\left\{\,
\begin{aligned}
\bs{\xi}&=\hat{\bs{\xi}}(\varphi,\Grad\varphi,\Grad^2\varphi)=\pards{\hat{\psi}(\varphi,\Grad\varphi,\Grad^2\varphi)}{(\Grad\varphi)},\\[4pt]
\bs{\mt{\Sigma}}&=\hat{\bs{\mt{\Sigma}}}(\varphi,\Grad\varphi,\Grad^2\varphi)=\pards{\hat{\psi}(\varphi,\Grad\varphi,\Grad^2\varphi)}{(\Grad^2\varphi)}.
\end{aligned}
\right.
\end{equation}
\item The internal microforce $\pi$ is given by a constitutive response function $\hat{\pi}$ that splits additively into a contribution derived from the response function $\hat{\psi}$ and a dissipative contribution that, in contrast to $\hat{\psi}$, $\hat{\bs{\xi}}$, and $\hat{\bs{\mt{\Sigma}}}$, depends on $\dot\varphi$ and must be consistent with a residual dissipation inequality:
\begin{equation}\label{eq:pi.constitutive.relation}
\left\{\,
\begin{aligned}
\pi=\hat{\pi}(\varphi,\Grad\varphi,\Grad^2\varphi,\dot\varphi)&=-\pards{\hat{\psi}(\varphi,\Grad\varphi,\Grad^2\varphi)}{\varphi}+\pi_{\textrm{dis}}(\varphi,\Grad\varphi,\Grad^2\varphi,\dot\varphi),\\[4pt]
\pi_{\textrm{dis}}(\varphi,\Grad\varphi,\Grad^2\varphi,\dot\varphi)\dot\varphi&\le0.
\end{aligned}
\right.
\end{equation}
\end{itemize}
In view of the constitutive restrictions \eqref{eq:free.energy.functional}--\eqref{eq:pi.constitutive.relation}, the response function for the free-energy density serves as a thermodynamic potential for the microstress, the hypermicrostress, and the equilibrium contribution to the internal microforce. A complete description of the response of a material belonging to the class in question thus consists of providing scalar-valued response functions $\hat{\psi}$ and $\pi_{\textrm{dis}}$. Whereas $\hat{\psi}$ depends only on $\varphi$, $\Grad\varphi$, and $\Grad^2\varphi$, $\pi_{\textrm{dis}}$ depends also on $\dot\varphi$. Moreover, $\pi_{\textrm{dis}}$ must satisfy the residual dissipation inequality \eqref{eq:pi.constitutive.relation}$_2$ for all choices of $\varphi$, $\Grad\varphi$, $\Grad^2\varphi$, and $\dot\varphi$.

\subsection{Nonconserved second-grade phase-field equation}

We now detail the logic of the derivation that leads us to define the Swift--Hohenberg equation as a particular case of the nonconserved second-grade phase-field equation. Thus, we here generalize the Swift--Hohenberg equation and fit this description within the second-grade phase-field framework we built in this section.

Using \eqref{eq:xi.Sigma.constitutive.relation} and \eqref{eq:pi.constitutive.relation} in the field equation \eqref{eq:field.equation}, we obtain an evolution equation
\begin{multline}\label{eq:2grade.phase.field.PDE}
-\pi_{\textrm{dis}}(\varphi,\Grad\varphi,\Grad^2\varphi,\dot\varphi)=\Div\bigg(\pards{\hat{\psi}(\varphi,\Grad\varphi,\Grad^2\varphi)}{(\Grad\varphi)}-\Div\bigg(\pards{\hat{\psi}(\varphi,\Grad\varphi,\Grad^2\varphi)}{(\Grad^2\varphi)}\bigg)\bigg)\\-\pards{\hat{\psi}(\varphi,\Grad\varphi,\Grad^2\varphi)}{\varphi}+\gamma,
\end{multline}
for the phase field.

We refer to \eqref{eq:2grade.phase.field.PDE} as a `nonconserved second-grade phase-field equation.'

\subsubsection{Swift--Hohenberg equation}

In particular, we choose $\hat{\psi}$ and $\pi_{\textrm{dis}}$ according to
\begin{equation}\label{eq:free.energy.dissipation.def}
\left\{\,
\begin{aligned}
\hat{\psi}(\varphi,\Grad\varphi,\Grad^2\varphi)&=f(\varphi)+\fr{1}{2}\lambda(\varphi^2 - 2\ell^2|\Grad\varphi|^2+\ell^4(\tr(\Grad^2\varphi))^2),\\[4pt]
\pi_{\textrm{dis}}(\varphi,\Grad\varphi,\Grad^2\varphi,\dot\varphi)&=-\beta\dot\varphi,
\end{aligned}
\right.
\end{equation}
where $f$ is a function of $\varphi$, and $\lambda>0$, $\ell>0$, $\beta>0$ are problem-specific-constants. Here, $f$ and $\lambda$ carry dimensions of energy per unit volume, $\ell$ carries dimensions of length, and $\beta$ carries dimensions of (dynamic) viscosity. Granted that $\hat{\psi}$ and $\pi_{\textrm{dis}}$ are as defined in \eqref{eq:free.energy.dissipation.def}, the thermodynamic restrictions \eqref{eq:xi.Sigma.constitutive.relation} and \eqref{eq:pi.constitutive.relation} yield
\begin{equation}\label{eq:xi.sigma.def}
\bs{\xi}=-2\lambda\ell^2\Grad\varphi,\qquad\bs{\mt{\Sigma}}=\lambda\ell^4(\triangle\varphi)\id,\qquad\pi=-f^\prime(\varphi)-\lambda\varphi-\beta\dot\varphi,
\end{equation}
where $\triangle=\Div\Grad$ denotes the Laplacian and a superposed prime denote differentiation with respect to $\varphi$. Using the particular constitutive relations \eqref{eq:xi.sigma.def} in the field equation \eqref{eq:field.equation}, we obtain the Swift--Hohenberg \cite{Swi77} equation
\begin{equation}\label{eq:sh}
\beta\dot\varphi=-\lambda(1+\ell^2\triangle)^2\varphi-f^\prime(\varphi)+\gamma.
\end{equation}

\subsection{Configurational fields}

Configurational forces are primitive entities that describe the motion of interfaces as well as the thermodynamics of their evolution. These forces are associated with the integrity of the material structure and the evolution of defects. Moreover, configurational forces expend power associated to the transfer of matter. Now, we recall the configurational balance presented by Fried \cite{Fri06b}, for a part $\prt$
\begin{equation}
\int\limits_{\dprt}\bs{C}\bs{n}\da+\int\limits_{\prt}(\bs{f}+\bs{e})\dv=\bs{0},
\end{equation}
which renders after localization the following pointwise version
\begin{equation}\label{eq:pointwise.configurational.balance}
\Div\bs{C}+\bs{f}+\bs{e}=\bs{0},
\end{equation}
where $\bs{C}$ is the configurational stress tensor (Eshelby stress tensor) while $\bs{f}$ and $\bs{e}$ are the internal and external forces.

The transfer of matter defines kinematic processes which characterize the power expended by the configurational forces. Thus, we first establish how configurational forces expend power in an immaterial migrating arbitrary part $\prt^\prime$, where $\bs{\upsilon}$ is the migrating boundary velocity defined on $\dprt^\prime$, letting $\bs{n}^\prime$ denote the outward unit normal. We consider that the migrating boundary $\dprt^\prime$ may exhibit a lack of smoothness along a curve $\edg^\prime$ with limiting normals $\{\bs{n}^+,\bs{n}^-\}$. Furthermore, the configurational traction $\bs{C}\bs{n}^\prime$ is assumed to be power conjugate to $\bs{\upsilon}$ on $\dprt^\prime$.

We now use the external virtual power \eqref{eq:external.virtual.power}, where $\gamma$, $\xis$, and $\tauc$ are conjugates to the virtual field $\chi$, while $\sigmas$ is conjugate to $\lpards{\chi}{n^\prime}$. Thus, we set as virtual fields the advective terms
\begin{equation}
\dot\varphi+\Grad\varphi\cdot\bs{\upsilon}\qquad\text{and}\qquad\dot{\overline{\left(\pards{\varphi}{n^\prime}\right)}}+\Grad\pards{\varphi}{n^\prime}\cdot\bs{\upsilon},
\end{equation}
to follow the motion of $\dprt^\prime$.

Next, with the surface microtraction \eqref{eq:surface.microtraction} and the hypermicrotraction \eqref{eq:hypermicrotraction}$_1$, consider the following identities
\begin{equation}\label{eq:advective.work}
\begin{split}
\xis(\dot\varphi+\Grad\varphi\cdot\bs{\upsilon})&=((\bs{\xi}-\Div\bs{\mt{\Sigma}})\cdot\bs{n}^\prime-\Divs(\bs{P}\mskip-2.5mu\bs{\mt{\Sigma}}\bs{n}^\prime))(\dot\varphi+\Grad\varphi\cdot\bs{\upsilon})\\
&=\xis\dot\varphi+\Grad\varphi\otimes(\bs{\xi}-\Div\bs{\mt{\Sigma}})\bs{n}^\prime\cdot\bs{\upsilon}+\bs{\mt{\Sigma}}\bs{n}^\prime\cdot\Grads(\Grad\varphi\cdot\bs{\upsilon})-\Divs(\bs{P}\mskip-2.5mu\bs{\mt{\Sigma}}\bs{n}^\prime\mskip+2.5mu\Grad\varphi\cdot\bs{\upsilon}),
\end{split}
\end{equation}
and
\begin{equation}
\sigmas\left(\pards{\dot\varphi}{n^\prime}+\Grad\pards{\varphi}{n^\prime}\cdot\bs{\upsilon}\right)=\sigmas\pards{\dot\varphi}{n^\prime}+\bs{n}^\prime\cdot\bs{\mt{\Sigma}}\bs{n}^\prime\left(\Grad\pards{\varphi}{n^\prime}\cdot\bs{\upsilon}\right).
\end{equation}
Now, bearing in mind that $\Grad\varphi\cdot\Grads\bs{\upsilon}=0$ together with the surface divergence theorem on a nonsmooth closed surface \eqref{eq:nonsmooth.divs.theo.closed.S}, we are led to the following external configurational power
\begin{multline}\label{eq:configurational.power}
\cl{W}_{\mathrm{ext}}(\prt^\prime)=\int\limits_{\prt^\prime}\gamma\dot\varphi\dv+\int\limits_{\dprt^\prime}\left(\bs{C}+\Grad\varphi\otimes(\bs{\xi}-\Div\bs{\mt{\Sigma}})+(\Grad^2\varphi)^{\trans}\bs{\mt{\Sigma}}\right)\bs{n}^\prime\cdot\bs{\upsilon}\da\\
+\int\limits_{\dprt^\prime}\left(\xis\dot\varphi+\sigmas\pards{\dot\varphi}{n^\prime}\right)\da+\int\limits_{\edg^\prime}\tauc\dot\varphi\ds.
\end{multline}
Since the nature of the motion of $\dprt^\prime$ involves only the normal component $\bs{\upsilon}\cdot\bs{n}^\prime$, the power must be indifferent to the tangential component of $\bs{\upsilon}$, implying
\begin{equation}\label{eq:configurational.stress.alpha}
\bs{C}+\Grad\varphi\otimes(\bs{\xi}-\Div\bs{\mt{\Sigma}})+(\Grad^2\varphi)^{\trans}\bs{\mt{\Sigma}}\eqqcolon\alpha\id,
\end{equation}
where $\alpha$ is a scalar field.

Thus, we can express the second integral of \eqref{eq:configurational.power} as
\begin{equation}
\int\limits_{\dprt^\prime}\alpha\mskip+2.5mu\bs{\upsilon}\cdot\bs{n}^\prime\da.
\end{equation}
Appealing to the free-energy imbalance \eqref{eq:partwise.imbalance} for a migrating arbitrary part $\prt^\prime$ with a velocity $\bs{\upsilon}$, we can state that
\begin{equation}
\dot{\overline{\int\limits_{\prt^\prime}\psi\dv}}=\int\limits_{\prt^\prime}\dot\psi\dv+\int\limits_{\dprt^\prime}\psi\bs{\upsilon}\cdot\bs{n}^\prime\da\le\int\limits_{\prt^\prime}\gamma\dot\varphi\dv+\int\limits_{\dprt^\prime}\alpha\bs{\upsilon}\cdot\bs{n}^\prime\da+\int\limits_{\dprt^\prime}\left(\xis\dot\varphi+\sigmas\pards{\dot\varphi}{n^\prime}\right)\da+\int\limits_{\edg^\prime}\tauc\dot\varphi\ds,
\end{equation}
leading to
\begin{equation}\label{eq:configurational.free.energy.imbalance}
\int\limits_{\prt^\prime}\dot\psi\dv\le\int\limits_{\prt^\prime}\gamma\dot\varphi\dv+\int\limits_{\dprt^\prime}(\alpha-\psi)\bs{\upsilon}\cdot\bs{n}^\prime\da+\int\limits_{\dprt^\prime}\left(\xis\dot\varphi+\sigmas\pards{\dot\varphi}{n^\prime}\right)\da+\int\limits_{\edg^\prime}\tauc\dot\varphi\ds,
\end{equation}
which implies that $\alpha=\psi$. Having $\alpha$ and with equation \eqref{eq:configurational.stress.alpha}, we determine the configurational stress. Invoking \eqref{eq:xi.Sigma.constitutive.relation}, \eqref{eq:pi.constitutive.relation}$_1$, and \eqref{eq:pointwise.configurational.balance}, we determine the internal and external configurational forces. Thus, the explicit form of the configurational stress tensor is
\begin{equation}
\bs{C}=\psi\id-\Grad\varphi\otimes(\bs{\xi}-\Div\bs{\mt{\Sigma}})-(\Grad^2\varphi)^{\trans}\bs{\mt{\Sigma}},
\end{equation}
while the internal and external configurational forces
\begin{equation}
\bs{f}=-\pi_{\textrm{dis}}\mskip+2.5mu\Grad\varphi\qquad\text{and}\qquad\bs{e}=-\gamma\mskip+2.5mu\Grad\varphi,
\end{equation}
respectively.

\section{Conserved second-grade phase field}
\label{conserved.case}

We here extend our theory to the case where the phase field represents the concentration of a conserved species with chemical potential $\mu$, flux $\bs{\jmath}$, and external rate of species production $s$, while continuing to restrict attention to isothermal processes. Following Gurtin's derivation of the Cahn--Hilliard equation \cite[\S3]{Gur96}, we therefore supplement the field equation \eqref{eq:field.equation} by a partwise species balance
\begin{equation}\label{eq:partwise.mass.balance}
\dot{\overline{\int\limits_{\prt}\varphi\dv}}=\int\limits_{\prt}s\dv-\int\limits_{\dprt}\bs{\jmath}\cdot\bs{n}\da.
\end{equation}
After localizing it, we obtain the pointwise version of the species balance
\begin{equation}\label{eq:pointwise.mass.balance}
\dot\varphi=s-\Div\bs{\jmath}
\end{equation}
Moreover, we augment the partwise free-energy imbalance \eqref{eq:partwise.imbalance} to account for the rate at which energy is transferred to $\prt$ due to species transport, yielding
\begin{equation}\label{eq:partwise.imbalance.conserved}
\dot{\overline{\int\limits_{\prt}\psi\dv}}\le\cl{W}_{\mathrm{ext}}(\prt)+\int\limits_{\prt}\mu s\dv-\int\limits_{\dprt}\mu\bs{\jmath}\cdot\bs{n}\da.
\end{equation}
Localizing \eqref{eq:partwise.imbalance.conserved} and using the field equation \eqref{eq:field.equation} and the pointwise species balance \eqref{eq:pointwise.mass.balance} to eliminate the external microforce $\gamma$ and rate of species production $s$, we arrive at the pointwise free-energy imbalance
\begin{equation}\label{eq:pointwise.imabalance.conserved}
\dot\psi+(\pi-\mu)\dot\varphi-\bs{\xi}\cdot\Grad\dot\varphi-\bs{\mt{\Sigma}}\colon\Grad^2\dot\varphi+\bs{\jmath}\cdot\Grad\mu\le0.
\end{equation}
Adding $\Grad^2\varphi$ and $\bs{\mt{\Sigma}}$ to the lists $(\varphi,\Grad\varphi,\mu,\Grad\mu)$ and $(\psi,\bs{\xi},\pi,\bs{\jmath})$ of independent and dependent constitutive variables considered by Gurtin \cite[\S3]{Gur96}, we find that the local inequality \eqref{eq:pointwise.imabalance.conserved} is satisfied in all processes if and only if:
\begin{itemize}
\item The free-energy density $\psi$ is given by a constitutive response function $\hat{\psi}$ that is independent of $\mu$ and $\Grad\mu$:
\begin{equation}
\psi=\hat{\psi}(\varphi,\Grad\varphi,\Grad^2\varphi).
\end{equation}
\item The microstress $\bs{\xi}$ and hypermicrostress $\bs{\mt{\Sigma}}$ are given by constitutive response functions $\hat{\bs{\xi}}$ and $\hat{\bs{\mt{\Sigma}}}$ that derive from the response function $\hat{\psi}$:
\begin{equation}\label{eq:xi.Sigma.constitutive.relation.conserved}
\left\{\,
\begin{aligned}
\bs{\xi}&=\hat{\bs{\xi}}(\varphi,\Grad\varphi,\Grad^2\varphi)=\pards{\hat{\psi}(\varphi,\Grad\varphi,\Grad^2\varphi)}{(\Grad\varphi)},\\[4pt]
\bs{\mt{\Sigma}}&=\hat{\bs{\mt{\Sigma}}}(\varphi,\Grad\varphi,\Grad^2\varphi)=\pards{\hat{\psi}(\varphi,\Grad\varphi,\Grad^2\varphi)}{(\Grad^2\varphi)}.
\end{aligned}
\right.
\end{equation}
\item The internal microforce $\pi$ is given by a constitutive response function $\hat{\pi}$ that differs from the chemical potential by a contribution derived from the response function $\hat{\psi}$:
\begin{equation}\label{eq:pi.constitutive.relation.conserved}
\pi=\hat{\pi}(\varphi,\Grad\varphi,\Grad^2\varphi,\mu)=\mu-\pards{\hat{\psi}(\varphi,\Grad\varphi,\Grad^2\varphi)}{\varphi}.
\end{equation}
\item Granted that the species flux $\bs{\jmath}$ depends smoothly on the gradient $\Grad\mu$ of the chemical potential $\mu$, it is given by a constitutive response function $\hat{\bs{\jmath}}$ of the form
\begin{equation}\label{eq:j.constitutive.relation.conserved}
\bs{\jmath}=\hat{\bs{\jmath}}(\varphi,\Grad\varphi,\Grad^2\varphi,\mu,\Grad\mu)=-\bs{M}(\varphi,\Grad\varphi,\Grad^2\varphi,\mu,\Grad\mu)\Grad\mu,
\end{equation}
where the mobility tensor $\bs{M}$ must obey the residual dissipation inequality
\begin{equation}\label{eq:dissipation.inequality.conserved}
\Grad\mu\cdot\bs{M}(\varphi,\Grad\varphi,\Grad^2\varphi,\mu,\Grad\mu)\Grad\mu\ge0
\end{equation}
for all choices of $\varphi$, $\Grad\varphi$, $\Grad^2\varphi$, $\mu$, and $\Grad\mu$.
\end{itemize}
In contrast to the theory previously developed for a phase field that is not a conserved species, a complete description of the response of a material belonging to the present class consists of providing a scalar-valued response function $\hat{\psi}$ and a tensor valued response function $\bs{M}$. Whereas $\hat{\psi}$ depends only on $\varphi$, $\Grad\varphi$, and $\Grad^2\varphi$, $\bs{M}$ may depend also on $\mu$ and $\Grad\mu$. Moreover, $\bs{M}$ must satisfy the residual dissipation inequality \eqref{eq:dissipation.inequality.conserved} for all choices of $\varphi$, $\Grad\varphi$, $\Grad^2\varphi$, $\mu$ and $\Grad\mu$.

\subsection{Conserved second-grade phase-field equation}

We now detail the logic of the derivation that leads us to define the phase-field crystal equation as a particular case of the conserved second-grade phase-field equation. Thus, we here generalize the phase-field crystal equation and fit this description within the conserved second-grade phase-field framework we built in this section.

Importantly, using \eqref{eq:xi.Sigma.constitutive.relation.conserved} and \eqref{eq:pi.constitutive.relation.conserved} in the field equation \eqref{eq:field.equation} generates the following expression
\begin{equation}\label{eq:mu.constitutive.relation.conserved}
\mu=\Div\bigg(\Div\bigg(\pards{\hat{\psi}(\varphi,\Grad\varphi,\Grad^2\varphi)}{(\Grad^2\varphi)}\bigg)-\pards{\hat{\psi}(\varphi,\Grad\varphi,\Grad^2\varphi)}{(\Grad\varphi)}\bigg)+\pards{\hat{\psi}(\varphi,\Grad\varphi,\Grad^2\varphi)}{\varphi}-\gamma,
\end{equation}
for the chemical potential which, in conjunction with the constitutive relation \eqref{eq:j.constitutive.relation.conserved} for the species flux and the pointwise species balance \eqref{eq:pointwise.mass.balance}, yields an evolution equation
\begin{equation}\label{eq:2grade.conserved.phase.field.PDE}
\dot\varphi=\Div(\bs{M}(\varphi,\Grad\varphi,\Grad^2\varphi,\mu,\Grad\mu)\Grad\mu)+s,
\end{equation}
for the phase field. Due to the dependence of the response function $\hat{\psi}$ on $\Grad^2\varphi$, $\mu$ as determined by \eqref{eq:mu.constitutive.relation.conserved} involves fourth-order spatial derivatives of $\varphi$ and \eqref{eq:2grade.conserved.phase.field.PDE} thus includes sixth-order spatial derivatives of $\varphi$.

In analogy to the comment immediately after the equation \eqref{eq:2grade.phase.field.PDE}, this suggests the possibility of referring to \eqref{eq:2grade.conserved.phase.field.PDE} with $\mu$ given by \eqref{eq:mu.constitutive.relation.conserved} as a `conserved second-grade phase-field equation.'

\subsubsection{Phase-field crystal equation}

Mimicking the assumptions leading from \eqref{eq:2grade.phase.field.PDE} to the Swift--Hohenberg equation \eqref{eq:sh}, we choose the definition of $\hat{\psi}$ to be
\begin{equation}\label{eq:free.energy.2gradePF}
\hat{\psi}(\varphi,\Grad\varphi,\Grad^2\varphi)=f(\varphi)+\fr{1}{2}\lambda(\varphi^2 - 2\ell^2|\Grad\varphi|^2+\ell^4(\tr(\Grad^2\varphi))^2),
\end{equation}
In addition, we stipulate that the mobility tensor depends at most on $\varphi$ and is isotropic, so that
\begin{equation}\label{eq:mobility}
\bs{M}(\varphi,\Grad\varphi,\Grad^2\varphi,\mu,\Grad\mu)=M\id,\qquad M>0.
\end{equation}
Then, \eqref{eq:xi.Sigma.constitutive.relation.conserved} and \eqref{eq:j.constitutive.relation.conserved} give
\begin{equation}\label{eq:xi.Sigma.j.def}
\bs{\xi}=-2\lambda\ell^2\Grad\varphi,\qquad\bs{\mt{\Sigma}}=\lambda\ell^4(\triangle\varphi)\id,\qquad\bs{\jmath}=-M\Grad\mu,
\end{equation}
while \eqref{eq:mu.constitutive.relation.conserved} specializes to
\begin{equation}\label{eq:mu.def}
\mu=f^\prime(\varphi)+\lambda(1+\ell^2\triangle)^2\varphi-\gamma.
\end{equation}
Combining \eqref{eq:xi.Sigma.j.def}$_3$ and \eqref{eq:mu.def} and using the resulting expression in the pointwise species balance \eqref{eq:2grade.conserved.phase.field.PDE}, we obtain the phase-field crystal equation
\begin{equation}\label{eq:pfc}
\dot\varphi=\Div\big(M\Grad(f^\prime(\varphi)+\lambda(1+\ell^2\triangle)^2\varphi-\gamma\big).
\end{equation}

\subsection{Configurational fields}

In accounting for species transport, species migration and the associated energy flow that occur in conjunction with the motion of the migrating part $\prt^\prime$, must be considered. The partwise species balance \eqref{eq:partwise.mass.balance} is rewritten as
\begin{equation}\label{eq:partwise.mass.balance.migrating}
\dot{\overline{\int\limits_{\prt^\prime}\varphi\dv}}-\int\limits_{\dprt^\prime}\varphi\bs{\upsilon}\cdot\bs{n}^\prime\da=\int\limits_{\prt^\prime}s\dv-\int\limits_{\dprt^\prime}\bs{\jmath}\cdot\bs{n}^\prime\da,
\end{equation}
where the free-energy imbalance for a migrating volume $\prt^\prime$ \eqref{eq:configurational.free.energy.imbalance} is specialized from \eqref{eq:partwise.imbalance.conserved}, becoming
\begin{multline}\label{eq:configurational.free.energy.imbalance.conserved}
\int\limits_{\prt^\prime}\dot\psi\dv-\int\limits_{\dprt^\prime}\mu\varphi\bs{\upsilon}\cdot\bs{n}^\prime\da\le\int\limits_{\prt^\prime}\gamma\dot\varphi\dv+\int\limits_{\prt^\prime}\mu s\dv+\int\limits_{\dprt^\prime}\left(\xis\dot\varphi+\sigmas\pards{\dot\varphi}{n^\prime}\right)\da\\
+\int\limits_{\dprt^\prime}(\alpha-\psi)\bs{\upsilon}\cdot\bs{n}^\prime\da-\int\limits_{\dprt^\prime}\mu\bs{\jmath}\cdot\bs{n}^\prime\da+\int\limits_{\edg^\prime}\tauc\dot\varphi\ds.
\end{multline}
Thus, the expressions \eqref{eq:advective.work} and \eqref{eq:configurational.free.energy.imbalance.conserved} yield the explicit form of the configurational stress tensor, the internal and external configurational forces, for the generalized conserved second-grade phase-field equation,
\begin{equation}
\bs{C}=(\psi-\mu\varphi)\id-\Grad\varphi\otimes(\bs{\xi}-\Div\bs{\mt{\Sigma}})-(\Grad^2\varphi)^{\trans}\bs{\mt{\Sigma}},\qquad\bs{f}=\varphi\mskip+2.5mu\Grad\mu,\qquad\text{and}\qquad\bs{e}=-\gamma\mskip+2.5mu\Grad\varphi.
\end{equation}

\section{Boundary conditions}
\label{boundary.conditions}

\subsection{Nonconserved second-grade phase field}

We here extend Fried \& Gurtin's \cite{Fri06a,Fri07} procedure, also expoited by Duda et al.\ \cite{Dud19} for the Cahn--Hilliard equation, to determine thermodynamically consistent boundary conditions by tailoring the surface balances of microforces and microtorques. In taking the surface $\srf$ in expressions \eqref{eq:surface.microforce.balance.smooth.open.S} and \eqref{eq:surface.microtorque.balance.smooth.open.S} (while including the terms relative to the internal-edge microtration) to the limit such that the surface coincides with the boundary, $\srf\subseteq\dbdy$, the surface $\xis$, surface-couple $\bsvarpis$, boundary-edge $\tauds$, and internal-edge $\tauc$ microtractions happen to represent external actions, ${\xis}_{\mathrm{env}}$, ${\bsvarpis}_{\mathrm{env}}$, ${\tauds}_{\mathrm{env}}$, and ${\tauc}_{\mathrm{env}}$, respectively, from the environment of $\bdy$. That is, in this limit the surface balances of microforces \eqref{eq:surface.microforce.balance.smooth.open.S} and the surface balances of microtorques \eqref{eq:surface.microtorque.balance.smooth.open.S} become
\begin{equation}\label{eq:boundary.balance.surface.microtractions}
\int\limits_{\srf}({\xis}_{\mathrm{env}}+\xis^\ast)\da+\int\limits_{\partial\srf}{\tauds}_{\mathrm{env}}\ds+\int\limits_{\edg}{\tauc}_{\mathrm{env}}\ds=0,\quad\forall\,\srf\subseteq\dbdy\quad\text{and}\quad t,
\end{equation}
and
\begin{equation}\label{eq:boundary.balance.surface.couple.microtractions}
\int\limits_{\srf}({\bsvarpis}_{\mathrm{env}}+\bsvarpis^\ast)\da+\int\limits_{\srf}({\xis}_{\mathrm{env}}+\xis^\ast)\bs{r}\da+\int\limits_{\partial\srf}{\tauds}_{\mathrm{env}}\bs{r}\ds+\int\limits_{\edg}{\tauc}_{\mathrm{env}}\bs{r}\ds=0,\quad\forall\,\srf\subseteq\dbdy\quad\text{and}\quad t,
\end{equation}
respectively.

With the \eqref{eq:surface.microtraction.opposite} and \eqref{eq:surface.microtraction}, we arrive at the following representation for the surface microtraction on $\srf^\ast$
\begin{equation}\label{eq:xis.ast}
\xis^\ast=-(\bs{\xi}-\Div\bs{\mt{\Sigma}})\cdot\bs{n}.
\end{equation}
Rewriting identity \eqref{eq:id.varpis.ast}, with \eqref{eq:surface.couple.microtraction.opposite} and \eqref{eq:surface.couple.microtraction}, we obtain the following representation for the surface-couple microtraction on $\srf^\ast$
\begin{equation}\label{eq:varpis.ast}
\bsvarpis^\ast=-\bs{\mt{\Sigma}}\bs{n}=-\Divs(\bs{r}\otimes\bs{P}\mskip-2.5mu\bs{\mt{\Sigma}}\bs{n})-(\bs{n}\otimes\bs{n})\bs{\mt{\Sigma}}\bs{n}+\bs{r}\mskip+2.5mu\Divs(\bs{P}\mskip-2.5mu\bs{\mt{\Sigma}}\bs{n}).
\end{equation}
From the surface balance of microtorques \eqref{eq:boundary.balance.surface.couple.microtractions}, with the representations \eqref{eq:xis.ast}, \eqref{eq:varpis.ast}, the surface divergence theorem on nonsmooth open surfaces \eqref{eq:nonsmooth.divs.theo.open.S}, and uncoupling \eqref{eq:boundary.balance.surface.couple.microtractions}, we are led to
\begin{equation}\label{eq:partwise.boundary.edge.microtraction.env}
\int\limits_{\dsrf}({\tauds}_{\mathrm{env}}-\bs{\nu}\cdot\bs{\mt{\Sigma}}\bs{n})\bs{r}\ds=0,\quad\forall\,\dsrf\subset\srf\quad\text{and}\quad t,
\end{equation}
\begin{equation}\label{eq:partwise.internal.edge.microtraction.env}
\int\limits_{\edg}({\tauc}_{\mathrm{env}}-\surp{\bs{\nu}\cdot\bs{\mt{\Sigma}}\bs{n}})\bs{r}\ds=0,\quad\forall\,\edg\subset\srf\quad\text{and}\quad t,
\end{equation}
and
\begin{equation}\label{eq:.partwise.surface.surface.couple.microtraction.env}
\int\limits_{\srf}\left(({\xis}_{\mathrm{env}}-(\bs{\xi}-\Div\bs{\mt{\Sigma}})\cdot\bs{n}+\Divs(\bs{P}\mskip-2.5mu\bs{\mt{\Sigma}}\bs{n}))+({\bsvarpis}_{\mathrm{env}}-(\bs{n}\otimes\bs{n})\bs{\mt{\Sigma}}\bs{n})\right)\bs{r}\da=0,\quad\forall\,\srf\subseteq\dbdy\quad\text{and}\quad t.
\end{equation}
Localizing expressions \eqref{eq:partwise.boundary.edge.microtraction.env} and \eqref{eq:partwise.internal.edge.microtraction.env}, we obtain the explicit representations of ${\tauds}_{\mathrm{env}}$ and ${\tauc}_{\mathrm{env}}$. Replacing ${\tauds}_{\mathrm{env}}$ and ${\tauc}_{\mathrm{env}}$ in the surface balance of microforces \eqref{eq:boundary.balance.surface.microtractions} and localizing it, we arrive at the representation of ${\xis}_{\mathrm{env}}$. Finally, with ${\xis}_{\mathrm{env}}$, ${\tauds}_{\mathrm{env}}$ and ${\tauc}_{\mathrm{env}}$ in \eqref{eq:.partwise.surface.surface.couple.microtraction.env} and localizing it, we are led to ${\bsvarpis}_{\mathrm{env}}$. Thus, the explicit form of the environmental microtractions are
\begin{equation}\label{eq:natural.prescribed}
\text{on }\srf_{\mathrm{nat}}\quad
\left\{
\begin{aligned}
{\xis}_{\mathrm{env}}&=(\bs{\xi}-\Div\bs{\mt{\Sigma}})\cdot\bs{n}-\Divs(\bs{P}\mskip-2.5mu\bs{\mt{\Sigma}}\bs{n}),\\[4pt]
{\bsvarpis}_{\mathrm{env}}&=(\bs{n}\otimes\bs{n})\bs{\mt{\Sigma}}\bs{n}\qquad\text{or}\qquad\sigma_{\mathrm{env}}=\bs{n}\cdot\bs{\mt{\Sigma}}\bs{n},\\[4pt]
{\tauc}_{\mathrm{env}}&=\surp{\bs{\nu}\cdot\bs{\mt{\Sigma}}\bs{n}},\\[4pt]
{\tauds}_{\mathrm{env}}&=\bs{\nu}\cdot\bs{\mt{\Sigma}}\bs{n}.
\end{aligned}
\right.
\end{equation}
The relations \eqref{eq:natural.prescribed}, represent the first set of suitable boundary conditions, where ${\xis}_{\mathrm{env}}$, $\sigma_{\mathrm{env}}$, ${\tauds}_{\mathrm{env}}$ and ${\tauc}_{\mathrm{env}}$ are given on $\srf$. In the variational context, these are natural boundary conditions on $\srf_{\mathrm{nat}}$.

Next, we require that the temporal increase in free energy of $\srf$ to be zero. This implies in particular that the power expended on $\srf$ be great or equal than zero. We express the partwise surface free-energy imbalance as
\begin{equation}\label{eq:free.energy.surf.env}
\cl{W}_{\mathrm{surf}}(\srf^\ast)+\cl{W}_{\mathrm{env}}(\srf)\ge0.
\end{equation}
The power expended on $\srf^\ast$ by the surface $\xis^\ast$ and surface-couple $\bsvarpis^\ast$ microtractions is given by
\begin{align}\label{eq:power.ast}
\cl{W}_{\mathrm{surf}}(\srf^\ast)&=\int\limits_{\srf}\left(\xis^\ast\dot{\varphi}+\bsvarpis^\ast\cdot\Grad\dot{\varphi}\right)\da\nonumber\\[4pt]
&=-\int\limits_{\srf}\left(\dot{\varphi}(\bs{\xi}-\Div\bs{\mt{\Sigma}})\cdot\bs{n}+\bs{\mt{\Sigma}}\bs{n}\cdot\Grad\dot{\varphi}\right)\da,\nonumber\\[4pt]
&=-\int\limits_{\srf}\left(((\bs{\xi}-\Div\bs{\mt{\Sigma}})\cdot\bs{n}-\Divs(\bs{P}\mskip-2.5mu\bs{\mt{\Sigma}}\bs{n}))\dot{\varphi}+\Divs(\dot{\varphi}\bs{P}\mskip-2.5mu\bs{\mt{\Sigma}}\bs{n})+\bs{n}\cdot\bs{\mt{\Sigma}}\bs{n}\pards{\dot{\varphi}}{n}\right)\da,\nonumber\\[4pt]
&=-\int_{\srf}\left(\xis\dot{\varphi}+\sigmas\pards{\dot{\varphi}}{n}\right)\da-\int_{\dsrf}\tauds\dot{\varphi}\ds-\int_{\edg}\tauc\dot{\varphi}\ds.
\end{align}
Thus, with \eqref{eq:power.ast} expression \eqref{eq:free.energy.surf.env} becomes
\begin{equation}\label{eq:free.energy.imbalance.surface}
-\int_{\srf}\left(\xis\dot{\varphi}+\sigmas\pards{\dot{\varphi}}{n}\right)\da-\int_{\dsrf}\tauds\dot{\varphi}\ds-\int_{\edg}\tauc\dot{\varphi}\ds+\cl{W}_{\mathrm{env}}(\srf)\ge0.
\end{equation}

For pasive environments, the quantities ${\xis}_{\mathrm{env}}$, $\sigma_{\mathrm{env}}$, ${\tauds}_{\mathrm{env}}$ and ${\tauc}_{\mathrm{env}}$ equal zero and so does the power expenditure exerted by the environment $\cl{W}_{\mathrm{env}}(\srf)$. However, we consider a more general setting and allow for non-homogeneous boundary conditions. Thus, we set the external power to be
\begin{equation}\label{eq:environment.nonconserved}
\cl{W}_{\mathrm{env}}(\srf) = \int_{\srf}\left({\xis}_{\mathrm{env}}\dot{\varphi}_{\mathrm{env}}+\sigma_{\mathrm{env}}\pards{\dot{\varphi}_{\mathrm{env}}}{n}\right)\da+\int_{\dsrf}{\tauds}_{\mathrm{env}}\dot{\varphi}_{\mathrm{env}}\ds+\int_{\edg}{\tauc}_{\mathrm{env}}\dot{\varphi}_{\mathrm{env}}\ds,
\end{equation}
where $\dot{\varphi}_{\mathrm{env}}$ and $\partial\dot{\varphi}_{\mathrm{env}}/\partial n$ are the limits of the time derivative of the environmental phase field and its normal derivative at $\srf$. Accounting for expression \eqref{eq:environment.nonconserved}, the partwise surface free-energy imbalance \eqref{eq:free.energy.imbalance.surface} renders
\begin{equation}\label{eq:free.energy.imbalance.surface.env}
\int_{\srf}\left({\xis}_{\mathrm{env}}\dot{\varphi}_{\mathrm{env}}-\xis\dot{\varphi}+\sigma_{\mathrm{env}}\pards{\dot{\varphi}_{\mathrm{env}}}{n}-\sigmas\pards{\dot{\varphi}}{n}\right)\da+\int_{\dsrf}({\tauds}_{\mathrm{env}}\dot{\varphi}_{\mathrm{env}}-\tauds\dot{\varphi})\ds+\int_{\edg}\left({\tauc}_{\mathrm{env}}\dot{\varphi}_{\mathrm{env}}-\tauc\dot{\varphi}\right)\ds\ge0.
\end{equation}
Since ${\xis}_{\mathrm{env}}=\xis$, $\sigma_{\mathrm{env}}=\sigmas$, ${\tauds}_{\mathrm{env}}=\tauds$, and ${\tauc}_{\mathrm{env}}=\tauc$, uncoupling the integral on different parts followed by localization, we obtain the following pointwise conditions
\begin{equation}\label{eq:active.condition}
(\dot{\varphi}_{\mathrm{env}}-\dot{\varphi}){\xis}_{\mathrm{env}}+\left(\pards{\dot{\varphi}_{\mathrm{env}}}{n}-\pards{\dot{\varphi}}{n}\right)\sigma_{\mathrm{env}}\ge0,\qquad(\dot{\varphi}_{\mathrm{env}}-\dot{\varphi}){\tauds}_{\mathrm{env}}\ge0,\qquad\text{and}\qquad(\dot{\varphi}_{\mathrm{env}}-\dot{\varphi}){\tauc}_{\mathrm{env}}\ge0.
\end{equation}
Expressions in \eqref{eq:active.condition} will serve us to design suitable boundary conditions for this continuum mechanical theory. In what follows, we propose some classes of boundary conditions. We focus on the uncouple case where
\begin{equation}\label{eq:separate.active.condition}
(\dot{\varphi}_{\mathrm{env}}-\dot{\varphi}){\xis}_{\mathrm{env}}\ge0,\qquad\left(\pards{\dot{\varphi}_{\mathrm{env}}}{n}-\pards{\dot{\varphi}}{n}\right)\sigma_{\mathrm{env}}\ge0,\qquad(\dot{\varphi}_{\mathrm{env}}-\dot{\varphi}){\tauds}_{\mathrm{env}}\ge0,\qquad\text{and}\qquad(\dot{\varphi}_{\mathrm{env}}-\dot{\varphi}){\tauc}_{\mathrm{env}}\ge0,
\end{equation}
which is sufficient but not necessary to guarantee the inequality direction of \eqref{eq:active.condition}.

The second possible set of boundary conditions is given by the trivial solution of \eqref{eq:separate.active.condition} when used as an equality, that is, the assignment
\begin{equation}\label{eq:essential.prescribed}
\text{on }\srf_{\mathrm{ess}}\quad
\left\{
\begin{aligned}
\dot{\varphi}_{\mathrm{env}}&=\dot{\varphi},\\[4pt]
\pards{\dot{\varphi}_{\mathrm{env}}}{n}&=\pards{\dot{\varphi}}{n}.
\end{aligned}
\right.
\end{equation}
These are essential boundary conditions on $\srf_{\mathrm{ess}}$. Note that, $\srf_{\mathrm{ess}}\cap\srf_{\mathrm{nat}}=\O$.

The third possible set of boundary conditions is given if $\dot{\varphi}_{\mathrm{env}}$ and $\partial\dot{\varphi}_{\mathrm{env}}/\partial n$ are prescribed, while ${\xis}_{\mathrm{env}}=\xis$, $\sigma_{\mathrm{env}}=\sigmas$, ${\tauds}_{\mathrm{env}}=\tauds$, and ${\tauc}_{\mathrm{env}}=\tauc$ are given by
\begin{equation}\label{eq:mixed.prescribed}
\text{on }\srf_{\mathrm{mix}}\quad
\left\{
\begin{aligned}
{\xis}_{\mathrm{env}}&=a\left(\dot{\varphi}_{\mathrm{env}}-\dot{\varphi}\right),\\[4pt]
\sigma_{\mathrm{env}}&=b\left(\pards{\dot{\varphi}_{\mathrm{env}}}{n}-\pards{\dot{\varphi}}{n}\right),\\[4pt]
{\tauds}_{\mathrm{env}}&=c\left(\dot{\varphi}_{\mathrm{env}}-\dot{\varphi}\right),\\[4pt]
{\tauc}_{\mathrm{env}}&=d\left(\dot{\varphi}_{\mathrm{env}}-\dot{\varphi}\right),
\end{aligned}
\right.
\end{equation}
with $a,b,c,d>0$. Combining ${\xis}_{\mathrm{env}}=\xis$, $\sigma_{\mathrm{env}}=\sigmas$, ${\tauds}_{\mathrm{env}}=\tauds$, and ${\tauc}_{\mathrm{env}}=\tauc$ from \eqref{eq:natural.prescribed} with \eqref{eq:mixed.prescribed}, we obtain the mixed boundary conditions on $\srf_{\mathrm{mix}}$.

\subsection{Boundary conditions for the classical Swift--Hohenberg equation}

As for the classical Swift--Hohenberg equation, although the essential boundary conditions \eqref{eq:essential.prescribed} remain the same, the natural \eqref{eq:natural.prescribed} and mixed \eqref{eq:mixed.prescribed} boundary conditions can be specialized. For the choice \eqref{eq:free.energy.dissipation.def}$_1$ of $\hat\psi$, \eqref{eq:xi.sigma.def}$_{1,2}$ yield $\bs{\xi}\cdot\bs{n}=-2\lambda\ell^2\partial\varphi/\partial n$, $\bs{n}\cdot\Div\bs{\mt{\Sigma}}=\lambda\ell^4\partial(\triangle\varphi)/\partial n$, $\bs{P}\mskip-2.5mu\bs{\mt{\Sigma}}\bs{n}=\lambda\ell^4\triangle\varphi\mskip+2.0mu\bs{P}\bs{n}=\bs{0}$, $\bs{n}\cdot\bs{\mt{\Sigma}}\bs{n}=\lambda\ell^4\triangle\varphi$, and $\bs{\nu}\cdot\bs{\mt{\Sigma}}\bs{n}=\lambda\ell^4\triangle\varphi\mskip+2.5mu\bs{\nu}\cdot\bs{n}=\bs{0}$. Thus, the natural boundary conditions \eqref{eq:natural.prescribed} read
\begin{equation}\label{eq:natural.prescribed.SH}
\text{on }\srf_{\mathrm{nat}}\quad
\left\{
\begin{aligned}
-\dfrac{1}{2\lambda\ell^2}{\xis}_{\mathrm{env}}&=\pards{\varphi}{n}+\dfrac{\ell^2}{2}\pards{(\triangle\varphi)}{n},\\[4pt]
\dfrac{1}{\lambda\ell^4}\sigma_{\mathrm{env}}&=\triangle\varphi,\\[4pt]
{\tauds}_{\mathrm{env}}&=0,\\[4pt]
{\tauc}_{\mathrm{env}}&=0,
\end{aligned}
\right.
\end{equation}
while the mixed boundary conditions \eqref{eq:mixed.prescribed}, when taking into account \eqref{eq:natural.prescribed.SH}, become
\begin{equation}\label{eq:mixed.prescribed.SH}
\text{on }\srf_{\mathrm{mix}}\quad
\left\{
\begin{aligned}
-\dfrac{1}{2\lambda\ell^2}a\left(\dot{\varphi}_{\mathrm{env}}-\dot{\varphi}\right)&=\pards{\varphi}{n}+\dfrac{\ell^2}{2}\pards{(\triangle\varphi)}{n},\\[4pt]
\dfrac{1}{\lambda\ell^4}b\left(\pards{\dot{\varphi}_{\mathrm{env}}}{n}-\pards{\dot{\varphi}}{n}\right)&=\triangle\varphi,\\[4pt]
{\tauds}_{\mathrm{env}}&=0,\\[4pt]
{\tauc}_{\mathrm{env}}&=0.
\end{aligned}
\right.
\end{equation}

\subsection{Conserved second-grade phase field}

In addition to the surface balances of microtractions \eqref{eq:boundary.balance.surface.microtractions} and microtorques \eqref{eq:boundary.balance.surface.couple.microtractions}, we supplement the system with the partwise surface species balance on $\srf$
\begin{equation}\label{eq:partwise.mass.balance.open.S}
\int_{\srf}(\jmath_{\mathrm{env}}-\bs{j}\cdot\bs{n})\da=0,
\end{equation}
which by localization renders
\begin{equation}\label{eq:pointwise.mass.balance.open.S}
\jmath_{\mathrm{env}}+\bs{j}\cdot\bs{n}=0.
\end{equation}
Here, $\jmath_{\mathrm{env}}$ represents the transfer of mass from the environment into $\srf$.

As we augmented the partwise free-energy imbalance \eqref{eq:partwise.imbalance} with the energy transfer rate to $\prt$ due to species transport to arrive at \eqref{eq:partwise.imbalance.conserved}, we augment \eqref{eq:free.energy.surf.env} with $\int_{\srf}\mu\bs{j}\cdot\bs{n}\da$ to obtain
\begin{equation}\label{eq:free.energy.surf.env.conserved}
\cl{W}_{\mathrm{surf}}(\srf^\ast)+\cl{W}_{\mathrm{env}}(\srf)+\int_{\srf}\mu\bs{j}\cdot\bs{n}\da\ge0.
\end{equation}
Analogously, \eqref{eq:environment.nonconserved} is augmented to become
\begin{equation}\label{eq:environment.conserved}
\cl{W}_{\mathrm{env}}(\srf) = \int_{\srf}\left({\xis}_{\mathrm{env}}\dot{\varphi}_{\mathrm{env}}+\sigma_{\mathrm{env}}\pards{\dot{\varphi}_{\mathrm{env}}}{n}+\jmath_{\mathrm{env}}\mu_{\mathrm{env}}\right)\da+\int_{\dsrf}{\tauds}_{\mathrm{env}}\dot{\varphi}_{\mathrm{env}}\ds+\int_{\edg}{\tauc}_{\mathrm{env}}\dot{\varphi}_{\mathrm{env}}\ds,
\end{equation}
where $\mu_{\mathrm{env}}$ is the limit of the environmental chemical potential at $\srf$. The surface free-energy imbalance, under the assumptions that led to \eqref{eq:free.energy.imbalance.surface.env}, from expression \eqref{eq:free.energy.surf.env.conserved} while accountig for \eqref{eq:environment.conserved}, we arrive at
\begin{multline}\label{eq:free.energy.imbalance.surface.env.conserved}
\int_{\srf}\left({\xis}_{\mathrm{env}}\dot{\varphi}_{\mathrm{env}}-\xis\dot{\varphi}+\sigma_{\mathrm{env}}\pards{\dot{\varphi}_{\mathrm{env}}}{n}-\sigmas\pards{\dot{\varphi}}{n}+\jmath_{\mathrm{env}}\mu_{\mathrm{env}}+\mu\bs{j}\cdot\bs{n}\right)\da\\
+\int_{\dsrf}({\tauds}_{\mathrm{env}}\dot{\varphi}_{\mathrm{env}}-\tauds\dot{\varphi})\ds+\int_{\edg}\left({\tauc}_{\mathrm{env}}\dot{\varphi}_{\mathrm{env}}-\tauc\dot{\varphi}\right)\ds\ge0.
\end{multline}
Following the same procedure that led to \eqref{eq:active.condition}$_1$ and taking into account the pointwise surface species balance \eqref{eq:pointwise.mass.balance.open.S}, we obtain the following additional inequality,
\begin{equation}\label{eq:active.condition.conserved}
\jmath_{\mathrm{env}}(\mu_{\mathrm{env}}-\mu)+(\dot{\varphi}_{\mathrm{env}}-\dot{\varphi}){\xis}_{\mathrm{env}}+\left(\pards{\dot{\varphi}_{\mathrm{env}}}{n}-\pards{\dot{\varphi}}{n}\right)\sigma_{\mathrm{env}}\ge0.
\end{equation}
Note that, \eqref{eq:active.condition}$_{2,3}$ remain the same. For the sake of simplicity, we opt to split and study \eqref{eq:active.condition.conserved} term by term as we did in \eqref{eq:separate.active.condition}. Thus, the only additional condition is
\begin{equation}\label{eq:separate.active.condition.conserved}
\jmath_{\mathrm{env}}(\mu_{\mathrm{env}}-\mu)\ge0.
\end{equation}

The first additional natural boundary condition arises from the pointwise surface species balance \eqref{eq:pointwise.mass.balance.open.S},
\begin{equation}\label{eq:natural.prescribed.conserved}
\bs{j}\cdot\bs{n}=-\jmath_{\mathrm{env}}\qquad\mathrm{on}\quad\srf_{\mathrm{nat}},
\end{equation}
where $\jmath_{\mathrm{env}}$ is prescribe and $\mu_{\mathrm{env}}$ is obtained from using \eqref{eq:separate.active.condition.conserved} as an equality.
Conversely, from \eqref{eq:separate.active.condition.conserved}, we obtain trivially the essential boundary condition,
\begin{equation}\label{eq:essential.prescribed.conserved}
\mu=\mu_{\mathrm{env}}\qquad\mathrm{on}\quad\srf_{\mathrm{ess}},
\end{equation}
where $\mu_{\mathrm{env}}$ is prescribed and $\jmath_{\mathrm{env}}$ is computed from the pointwise surface species balance \eqref{eq:pointwise.mass.balance.open.S}.

A mixed boundary condition is obtained by prescribing $\mu_{\mathrm{env}}$ and evaluating the transfer of mass from the environment with
\begin{equation}\label{eq:mixed.prescribed.conserved}
\jmath_{\mathrm{env}}=c\left(\mu_{\mathrm{env}}-\mu\right)\qquad\mathrm{on}\quad\srf_{\mathrm{mix}},
\end{equation}
with $c>0$. Finally, accounting for the pointwise surface species balance \eqref{eq:pointwise.mass.balance.open.S}, the expression \eqref{eq:natural.prescribed.conserved} is rendered as
\begin{equation}\label{eq:mixed.prescribed.conserved}
\bs{j}\cdot\bs{n}=c\left(\mu_{\mathrm{env}}-\mu\right)\qquad\mathrm{on}\quad\srf_{\mathrm{mix}}.
\end{equation}

\subsection{Boundary conditions for the classical phase-field crystal equation}

For the choice \eqref{eq:mobility} of $\bs{M}$, \eqref{eq:xi.Sigma.j.def}$_3$ yields $\bs{j}\cdot\bs{n}=-M(\varphi)\partial\mu/\partial n$ where $\mu$ is given by \eqref{eq:mu.def}. Thus, the particular versions of the natural, essential, and mixed boundary conditions \eqref{eq:natural.prescribed.conserved}, \eqref{eq:essential.prescribed.conserved}, and \eqref{eq:mixed.prescribed.conserved} that pertain to the classical phase-field crystal equation \eqref{eq:2grade.conserved.phase.field.PDE} are
\begin{equation}\label{eq:neumann.pfc}
M(\varphi)\pards{(\lambda(1+\ell^2\triangle)^2\varphi+f^\prime(\varphi))}{n}=-\jmath_{\mathrm{env}}
\end{equation}
on $\srf_{\textrm{nat}}$,
\begin{equation}\label{eq:dirichlet.pfc}
\lambda(1+\ell^2\triangle)^2\varphi+f^\prime(\varphi)=\mu_{\mathrm{env}},
\end{equation}
on $\srf_{\textrm{ess}}$,
\begin{equation}\label{eq:mixed.pfc}
M(\varphi)\pards{(\lambda(1+\ell^2\triangle)^2\varphi+f^\prime(\varphi))}{n}=c\left(\mu_{\mathrm{env}}-\lambda(1+\ell^2\triangle)^2\varphi+f^\prime(\varphi)\right)
\end{equation}
on $\srf_{\textrm{mix}}$, respectively.

\section{Summary}
\label{conclusions}

The continuum mechanical theory we introduce in this paper allows us to describe the underlying mechanical interactions which yield popular phase-field models such as the Swift--Hohenberg and phase-field crystal equations. We also recognize and account for the lack of smoothness in arbitrary parts which renders additional interations between different parts. In considering these interactions, we generalize these phase-field models. We summarize the results as follows.

\begin{rmk}[\textbf{\textit{Summary of results independent of constitutive relations}}]
The fundamental fields and the field equations are\\
\begin{enumerate}[label=(\roman*),font=\itshape,leftmargin=*]
\item \emph{Surface microtraction on $\srf$:}
\begin{equation*}
\xis=(\bs{\xi} - \Div\bs{\mt{\Sigma}})\cdot\bs{n} - \Divs(\bs{P}\mskip-2.5mu\bs{\mt{\Sigma}}\bs{n}).
\end{equation*}
\item \emph{Surface microtraction on the opposite side of $\srf$:}
\begin{equation*}
-\xis^\ast=(\bs{\xi} - \Div\bs{\mt{\Sigma}})\cdot\bs{n}.
\end{equation*}
\item \emph{Surface-couple microtraction on $\srf$:}
\begin{equation*}
\bsvarpis=(\bs{n}\otimes\bs{n})\bs{\mt{\Sigma}}\bs{n}.
\end{equation*}
\item \emph{Surface-couple microtraction on the opposite side of $\srf$:}
\begin{equation*}
-\bsvarpis^\ast=\bs{\mt{\Sigma}}\bs{n}.
\end{equation*}
\item \emph{Boundary-edge microtraction on $\dprt$:}
\begin{equation*}
\tauds=\bs{\nu}\cdot\bs{\mt{\Sigma}}\bs{n}.
\end{equation*}
\item \emph{Internal-edge microtraction on $\edg$:}
\begin{equation*}
\tauc=\surp{\bs{\nu}\cdot\bs{\mt{\Sigma}}\bs{n}}.
\end{equation*}
\end{enumerate}
\begin{enumerate}[label=(\roman*),font=\itshape,leftmargin=*,resume]
\item \emph{Hypermicrotraction (not a fundamental field) on $\srf$:}
\begin{equation*}
\sigmas\coloneqq\bs{n}\cdot\bs{\mt{\Sigma}}\bs{n}\quad\Longrightarrow\quad\bsvarpis=\sigmas\bs{n}.
\end{equation*}
\end{enumerate}
\begin{enumerate}[label=(\roman*),font=\itshape,leftmargin=*,resume]
\item \emph{Balance of microforces on a part $\prt$:}
\begin{equation*}
\Div\bs{\zeta}+\pi+\gamma=0.
\end{equation*}
\item \emph{Balance of microtroques on a part $\prt$:}
\begin{equation*}
\bs{\zeta}=\bs{\xi}-\Div\bs{\mt{\Sigma}}.
\end{equation*}
\item \emph{Species balance:}
\begin{equation*}
\dot\varphi=-\Div\bs{\jmath}+s.
\end{equation*}
\item \emph{The virtual power balance on a part $\prt$:}
\begin{gather*}
\underbrace{\int\limits_{\prt}(-\pi\chi+\bs{\xi}\cdot\Grad\chi+\bs{\mt{\Sigma}}\colon\Grad^2\chi)\dv}_{\cl{V}_{\text{int}}(\prt;\chi)}=\underbrace{\int\limits_{\prt}\gamma\chi\dv+\int\limits_{\dprt}\Big(\xis\chi+\sigmas\pards{\chi}{n}\Big)\da+\int\limits_{\edg}\tauc\chi\ds}_{\cl{V}_{\text{ext}}(\prt;\chi)},\\
\forall\,\prt\subseteq\bdy\quad\text{and}\quad t,
\end{gather*}
where $\chi$ is a smooth and admissible scalar virtual field and $\partial{\chi}/\partial{n}$ its normal derivative.
\item \emph{Configurational balance:}
\begin{equation*}
\Div\bs{C}+\bs{f}+\bs{e}=\bs{0}.
\end{equation*}
\item \emph{Configurational fields, for a nonconserved phase field:}
\begin{equation*}
\bs{C}=\psi\id-\Grad\varphi\otimes(\bs{\xi}-\Div\bs{\mt{\Sigma}})-(\Grad^2\varphi)^{\trans}\bs{\mt{\Sigma}},\qquad\bs{f}=-\pi_{\textrm{dis}}\mskip+2.5mu\Grad\varphi,\qquad\text{and}\qquad\bs{e}=-\gamma\mskip+2.5mu\Grad\varphi.
\end{equation*}
\item \emph{Configurational fields, for a conserved phase field:}
\begin{equation*}
\bs{C}=(\psi-\mu\varphi)\id-\Grad\varphi\otimes(\bs{\xi}-\Div\bs{\mt{\Sigma}})-(\Grad^2\varphi)^{\trans}\bs{\mt{\Sigma}},\qquad\bs{f}=\varphi\mskip+2.5mu\Grad\mu,\qquad\text{and}\qquad\bs{e}=-\gamma\mskip+2.5mu\Grad\varphi.
\end{equation*}
\item Natural boundary conditions:
\begin{equation*}
\text{on }\srf_{\mathrm{nat}}\quad
\left\{
\begin{aligned}
{\xis}_{\mathrm{env}}&=(\bs{\xi}-\Div\bs{\mt{\Sigma}})\cdot\bs{n}-\Divs(\bs{P}\mskip-2.5mu\bs{\mt{\Sigma}}\bs{n}),\\[4pt]
{\bsvarpis}_{\mathrm{env}}&=(\bs{n}\otimes\bs{n})\bs{\mt{\Sigma}}\bs{n}\qquad\text{or}\qquad\sigma_{\mathrm{env}}=\bs{n}\cdot\bs{\mt{\Sigma}}\bs{n},\\[4pt]
{\tauds}_{\mathrm{env}}&=\bs{\nu}\cdot\bs{\mt{\Sigma}}\bs{n},\\[4pt]
{\tauc}_{\mathrm{env}}&=\surp{\bs{\nu}\cdot\bs{\mt{\Sigma}}\bs{n}}.
\end{aligned}
\right.
\end{equation*}
\item Essential boundary conditions:
\begin{equation*}
\text{on }\srf_{\mathrm{ess}}\quad
\left\{
\begin{aligned}
\dot{\varphi}_{\mathrm{env}}&=\dot{\varphi},\\[4pt]
\pards{\dot{\varphi}_{\mathrm{env}}}{n}&=\pards{\dot{\varphi}}{n}.
\end{aligned}
\right.
\end{equation*}
\item Mixed boundary conditions:
\begin{equation*}
\text{on }\srf_{\mathrm{mix}}\quad
\left\{
\begin{aligned}
(\bs{\xi}-\Div\bs{\mt{\Sigma}})\cdot\bs{n}-\Divs(\bs{P}\mskip-2.5mu\bs{\mt{\Sigma}}\bs{n})&=a\left(\dot{\varphi}_{\mathrm{env}}-\dot{\varphi}\right),\\[4pt]
\bs{n}\cdot\bs{\mt{\Sigma}}\bs{n}&=b\left(\pards{\dot{\varphi}_{\mathrm{env}}}{n}-\pards{\dot{\varphi}}{n}\right),\\[4pt]
\bs{\nu}\cdot\bs{\mt{\Sigma}}\bs{n}&=c\left(\dot{\varphi}_{\mathrm{env}}-\dot{\varphi}\right),\\[4pt]
\surp{\bs{\nu}\cdot\bs{\mt{\Sigma}}\bs{n}}&=d\left(\dot{\varphi}_{\mathrm{env}}-\dot{\varphi}\right),
\end{aligned}
\right.
\end{equation*}
with $a,b,c,d>0$.
\item Natural boundary conditions for conserved species:
\begin{equation*}
\bs{j}\cdot\bs{n}=-\jmath_{\mathrm{env}}\qquad\mathrm{on}\quad\srf_{\mathrm{nat}}.
\end{equation*}
\item Essential boundary conditions for conserved species:
\begin{equation*}
\mu=\mu_{\mathrm{env}}\qquad\mathrm{on}\quad\srf_{\mathrm{ess}}.
\end{equation*}
\item Mixed boundary conditions for conserved species:
\begin{equation*}
\bs{j}\cdot\bs{n}=c\left(\mu_{\mathrm{env}}-\mu\right)\qquad\mathrm{on}\quad\srf_{\mathrm{mix}}.
\end{equation*}
\end{enumerate}
\qed
\end{rmk}

\begin{rmk}[\textbf{\textit{Constitutive relations}}]
For a nonconserved phase field, \emph{Internal microforce, microstress, and hypermicrostress:}
\begin{equation*}
\left\{\,
\begin{aligned}
\pi&=\hat{\pi}(\varphi,\Grad\varphi,\Grad^2\varphi,\dot\varphi)=-\pards{\hat{\psi}(\varphi,\Grad\varphi,\Grad^2\varphi)}{\varphi}+\pi_{\textrm{dis}}(\varphi,\Grad\varphi,\Grad^2\varphi,\dot\varphi),\\[4pt]
\bs{\xi}&=\hat{\bs{\xi}}(\varphi,\Grad\varphi,\Grad^2\varphi)=\pards{\hat{\psi}(\varphi,\Grad\varphi,\Grad^2\varphi)}{(\Grad\varphi)},\\[4pt]
\bs{\mt{\Sigma}}&=\hat{\bs{\mt{\Sigma}}}(\varphi,\Grad\varphi,\Grad^2\varphi)=\pards{\hat{\psi}(\varphi,\Grad\varphi,\Grad^2\varphi)}{(\Grad^2\varphi)}.
\end{aligned}
\right.
\end{equation*}
For a conserved phase field, \emph{Chemical potential, mass flux, internal microforce, microstress, and hypermicrostress:}
\begin{equation*}
\left\{\,
\begin{aligned}
\mu&=\Div\bigg(\Div\bigg(\pards{\hat{\psi}(\varphi,\Grad\varphi,\Grad^2\varphi)}{(\Grad^2\varphi)}\bigg)-\pards{\hat{\psi}(\varphi,\Grad\varphi,\Grad^2\varphi)}{(\Grad\varphi)}\bigg)+\pards{\hat{\psi}(\varphi,\Grad\varphi,\Grad^2\varphi)}{\varphi},\\[4pt]
\bs{\jmath}&=\hat{\bs{\jmath}}(\varphi,\Grad\varphi,\Grad^2\varphi,\mu,\Grad\mu)=-\bs{M}(\varphi,\Grad\varphi,\Grad^2\varphi,\mu,\Grad\mu)\Grad\mu,\\[4pt]
\pi&=\hat{\pi}(\varphi,\Grad\varphi,\Grad^2\varphi,\mu)=\mu-\pards{\hat{\psi}(\varphi,\Grad\varphi,\Grad^2\varphi)}{\varphi},\\[4pt]
\bs{\xi}&=\hat{\bs{\xi}}(\varphi,\Grad\varphi,\Grad^2\varphi)=\pards{\hat{\psi}(\varphi,\Grad\varphi,\Grad^2\varphi)}{(\Grad\varphi)},\\[4pt]
\bs{\mt{\Sigma}}&=\hat{\bs{\mt{\Sigma}}}(\varphi,\Grad\varphi,\Grad^2\varphi)=\pards{\hat{\psi}(\varphi,\Grad\varphi,\Grad^2\varphi)}{(\Grad^2\varphi)}.
\end{aligned}
\right.
\end{equation*}
\qed
\end{rmk}


\section{Acknowledgments}

We are indebted to Professor Eliot Fried. We had many exhaustive discussions in which he gave us valuable ideas, constructive comments, and encouragement. The first author would also like to thank his little brother, Daniel, for drawing the pictures used in this work. This publication was made possible in part by the CSIRO Professorial Chair in Computational Geoscience at Curtin University and the Deep Earth Imaging Enterprise Future Science Platforms of the Commonwealth Scientific Industrial Research Organisation, CSIRO, of Australia. The European Union's Horizon 2020 Research and Innovation Program of the Marie Sk\l{}odowska-Curie grant agreement No. 777778, and the Mega-grant of the Russian Federation Government (N 14.Y26.31.0013) provided additional support. Lastly, we acknowledge the support provided at Curtin University by The Institute for Geoscience Research (TIGeR) and by the Curtin Institute for Computation.


\appendix

\section{Identities}
\label{identities}

Let $\bs{u}$ be a sufficiently smooth displacement field inducing the motion $\srf\mapsto\srf_{\epsilon}$. Parameterizing the evolution of $\srf$, and letting
\begin{equation}\label{eq:parameterization.motion}
\bs{y}_\epsilon\coloneqq\bs{x}+\epsilon\bs{u}(\bs{x}),\quad\epsilon\in[0,1],
\end{equation}
with $\bs{x}\in\bdy$ and $\srf_\epsilon\coloneqq\bs{y}_\epsilon(\srf)$, we define the deformation gradient
\begin{equation}
\bs{F}\coloneqq\Grad\bs{y}_\epsilon\vert_{\epsilon=1}\qquad\text{and}\qquad\bs{F}_\epsilon\coloneqq\Grad\bs{y}_\epsilon,
\end{equation}
or equivalently
\begin{equation}
\bs{F}=\id+\Grad\bs{u}\qquad\text{and}\qquad\bs{F}_\epsilon=(1-\epsilon)\id+\epsilon\bs{F}.
\end{equation}
Quantities with the subscript $\epsilon$ refer to the deformed configuration, whereas quantities without the subscript live on the reference or undeformed configuration. The point $\bs{x}$ represents a point in the undeformed configuration $\srf$, whereas $\bs{y}_\epsilon$ represents its mapping onto the deformed configuration $\srf_\epsilon$. The function $\bs{y}_\epsilon$ is a---one-parameter family---parameterization, which linearizes the mapping $\bs{x}\mapsto\bs{y}_\epsilon$ along the displacement $\bs{u}$. Thus, $\bs{F}$ is the total deformation gradient, while $\bs{F}_\epsilon$ is the parameterized deformation gradient. Finally, let $J\coloneqq\det\bs{F}$.

Gurtin et al.\ \cite[(equations (6.15) and (8.4))]{Gur10} presents deformations laws for the unit tangent and unit normal vectors of a surface undergoing deformation from $\srf$ to $\srf_\epsilon$. These vectors preserve their norm under the following transformations
\begin{subequations}
\begin{align}
\bs{t}_\epsilon&=\dfrac{1}{|\bs{F}_\epsilon\bs{t}|}\bs{F}_\epsilon\bs{t},\\
\bs{n}_\epsilon&=\dfrac{1}{|\bs{F}^{-\trans}_\epsilon\bs{n}|}\bs{F}^{-\trans}_\epsilon\bs{n},
\end{align}
\end{subequations}
while the elements of length $\!\ds$, area $\!\da$, and volume $\!\dv$, \cite[(equations (7.14), (8.20), and (8.9))]{Gur10} transform with
\begin{subequations}\label{eq:d}
\begin{align}
\ds_\epsilon&=|\bs{F}\bs{t}|\ds+o(\!\ds)\footnotemark,\label{eq:ds}\\
\da_\epsilon&=J|\bs{F}^{-\trans}\bs{n}|\da+o(\!\da),\label{eq:da}\\
\dv_\epsilon&=J\dv+o(\!\dv).\label{eq:dv}
\end{align}
\end{subequations}
\footnotetext{$o(\ell)$ has the usual connotation $\lim\limits_{\ell\to\infty} o(\ell)\ell^{-1}=0$ as well as $\cl{O}(\ell)$ is defined such that $\lim\limits_{\ell\to\infty} \cl{O}(\ell)\ell^{-1}=c$, with $c>0$.}

Additionally, it is useful to express the gradient of $\bs{u}$ and its projection on $\srf$ as
\begin{equation}
\Grad\bs{u}=\pards{\bs{u}}{n}\otimes\bs{n}+\Grads\bs{u}\qquad\text{and}\qquad\Grads\bs{u}=(\Grad\bs{u})\bs{P},
\end{equation}
where
\begin{equation}
\pards{\bs{u}}{s}=(\Grad\bs{u})\bs{t},\qquad\pards{\bs{u}}{n}=(\Grad\bs{u})\bs{n},\qquad\text{and}\qquad\pards{\bs{u}}{\nu}=(\Grad\bs{u})\bs{\nu}.
\end{equation}
It is also convenient to express a vector $\bs{a}$ as $\bs{a}=(\bs{a}\cdot\bs{t})\bs{t}+(\bs{a}\cdot\bs{n})\bs{n}+(\bs{a}\cdot\bs{\nu})\bs{\nu}$. Finally, we denote the curvature tensor by $\bs{L}\coloneqq-\Grads\bs{n}$ while the mean curvature by $L\coloneqq-\frac{1}{2}\Divs\bs{n}$.

We here aim at computing the following derivatives
\begin{equation}\label{eq:rates.tensors}
\begin{split}
\dd{\bs{n}_\epsilon}{\epsilon}\Big\vert_{\epsilon=0},\qquad\dd{\bs{t}_\epsilon}{\epsilon}\Big\vert_{\epsilon=0},\qquad\dd{\bs{\nu}_\epsilon}{\epsilon}\Big\vert_{\epsilon=0},\qquad\dd{\bs{L}_\epsilon}{\epsilon}\Big\vert_{\epsilon=0},\\
\dd{(|\bs{F}_\epsilon\bs{t}_\epsilon|)}{\epsilon}\Big\vert_{\epsilon=0},\qquad\dd{(J_\epsilon|\bs{F}^{-\trans}_\epsilon\bs{n}_\epsilon|)}{\epsilon}\Big\vert_{\epsilon=0},\qquad\text{and}\qquad\dd{J_\epsilon}{\epsilon}\Big\vert_{\epsilon=0}.
\end{split}
\end{equation}
The last three quantities \eqref{eq:rates.tensors}$_{5,6,7}$ represent, respectively, the rate of change of elements of length, area, and volume.

In order to compute the inverse of the deformation gradient, by Neumann series \cite{Ste98}, given the invertible second-order tensor $\bs{A}$ and the second-order tensor $\bs{B}$, we define the following identity
\begin{align}
(\bs{A}+\epsilon\bs{B})^{-1}&=(\bs{A}(\id+\epsilon\bs{A}^{-1}\bs{B}))^{-1}\nonumber\\
&=(\id+\epsilon\bs{A}^{-1}\bs{B})^{-1}\bs{A}^{-1}\nonumber\\
&=(\id-\epsilon\bs{A}^{-1}\bs{B}+\epsilon^2\bs{A}^{-1}\bs{B}\bs{A}^{-1}\bs{B}-\ldots)\bs{A}^{-1}\nonumber\\
&=(\bs{A}^{-1}-\epsilon\bs{A}^{-1}\bs{B}\bs{A}^{-1})+\cl{O}(\epsilon^2).
\end{align}
and letting $\bs{A}\coloneqq(1-\epsilon)\id$ and $\bs{B}\coloneqq\bs{F}$, we arrive at
\begin{align}
\bs{F}_\epsilon^{-1}&=((1-\epsilon)\id+\epsilon\bs{F})^{-1}\nonumber\\
&=(1-\epsilon)^{-1}\id-\epsilon(1-\epsilon)^{-1}\id\bs{F}(1-\epsilon)^{-1}\id+\cl{O}(\epsilon^2)\nonumber\\
&=(1-\epsilon)^{-1}\id-\epsilon(1-\epsilon)^{-2}\bs{F}+\cl{O}(\epsilon^2).
\end{align}
When it comes to analyzing $\bs{A}^{-1}$, with $\bs{A}=(1-\epsilon)\id$, we restrict our analysis to the range $[0,1[$ without loss of generality.

To determine the rate of change of the unit normal with respect to $\epsilon$ in \eqref{eq:rates.tensors}$_1$ when a surface undergoes deformation from $\srf\mapsto\srf_\epsilon$, we establish the following identities. The rate of change of the inverse of the deformation gradient with respect to $\epsilon$ is
\begin{align}
\dd{(\bs{F}^{-1}_\epsilon)}{\epsilon}\Big\vert_{\epsilon=0}&=\dfrac{1}{(1-\epsilon)^2}\id\Big\vert_{\epsilon=0}-\dfrac{1+\epsilon}{(1-\epsilon)^3}\bs{F}\Big\vert_{\epsilon=0}+\dd{\cl{O}(\epsilon^2)}{\epsilon}\Big\vert_{\epsilon=0},\nonumber\\[4pt]
&=\id-\bs{F}=-\Grad\bs{u},
\end{align}
while the unit normal presents the following rate of change
\begin{equation}\label{eq:rate.normal.no.preserving}
\dd{(\bs{F}^{-\trans}_\epsilon\bs{n})}{\epsilon}\Big\vert_{\epsilon=0}=-(\Grad\bs{u})^{\trans}\bs{n}.
\end{equation}
Although it does not preserve its norm. Thus, to  determine the rate of change of inverse of the norm of the normal, $|\bs{F}^{-\trans}_\epsilon\bs{n}|^{-1}$, we first compute
\begin{align}
\dd{(\bs{F}^{-1}_\epsilon\bs{F}^{-\trans}_\epsilon)}{\epsilon}\Big\vert_{\epsilon=0}&=\dd{}{\epsilon}\left[\dfrac{1}{(1-\epsilon)^2}\id+\dfrac{\epsilon^2}{(1-\epsilon)^4}\bs{F}\bs{F}^{\trans}-\dfrac{\epsilon}{(1-\epsilon)^3}(\bs{F}^{\trans}+\bs{F})\right]_{\epsilon=0}+\dd{\cl{O}(\epsilon^2)}{\epsilon}\Big\vert_{\epsilon=0},\nonumber\\[4pt]
&=\left[\dfrac{2}{(1-\epsilon)^3}\id+2\epsilon\dfrac{1+\epsilon}{(1-\epsilon)^5}\bs{F}\bs{F}^{\trans}-\dfrac{1+2\epsilon}{(1-\epsilon)^4}(\bs{F}^{\trans}+\bs{F})\right]_{\epsilon=0}+\dd{\cl{O}(\epsilon^2)}{\epsilon}\Big\vert_{\epsilon=0},\nonumber\\[4pt]
&=2\id-(\bs{F}^{\trans}+\bs{F})=-2\mskip+2.5mu\sym(\Grad\bs{u}),
\end{align}
to compute the rate of change of the norm, $|\bs{F}^{-\trans}_\epsilon\bs{n}|=\sqrt{\bs{F}^{-1}_\epsilon\bs{F}^{-\trans}_\epsilon\colon\bs{n}\otimes\bs{n}}$,
\begin{equation}
\dd{(\sqrt{\bs{F}^{-1}_\epsilon\bs{F}^{-\trans}_\epsilon\colon\bs{n}\otimes\bs{n}})}{\epsilon}\Big\vert_{\epsilon=0}=-\sym(\Grad\bs{u})\colon\bs{n}\otimes\bs{n},
\end{equation}
to arrive at
\begin{equation}\label{eq:rate.inv.norm.normal}
\dd{(|\bs{F}^{-\trans}_\epsilon\bs{n}|^{-1})}{\epsilon}\Big\vert_{\epsilon=0}=\sym(\Grad\bs{u})\colon\bs{n}\otimes\bs{n}=\Grad\bs{u}\colon\bs{n}\otimes\bs{n}.
\end{equation}
Now, with expressions \eqref{eq:rate.normal.no.preserving} and \eqref{eq:rate.inv.norm.normal}, we obtain the rate of change of the unit normal which preserves the norm
\begin{equation}\label{eq:rate.normal}
\dd{\bs{n}_\epsilon}{\epsilon}\Big\vert_{\epsilon=0}=(\Grad\bs{u}\colon\bs{n}\otimes\bs{n})\bs{n}-(\Grad\bs{u})^{\trans}\bs{n}=-(\Grads\bs{u})^{\trans}\bs{n}.
\end{equation}

To determine the rate of change of the unit tangent with respect to $\epsilon$ in \eqref{eq:rates.tensors}$_2$ when a surface undergoes deformation from $\srf\mapsto\srf_\epsilon$, we establish the following identities. The unit tangent presents the following rate of change
\begin{equation}\label{eq:rate.tangent.no.preserving}
\dd{(\bs{F}_\epsilon\bs{t})}{\epsilon}\Big\vert_{\epsilon=0}=(\Grad\bs{u})\bs{t}.
\end{equation}
Again, this rate does not preserve its norm. Thus, to determine the rate of change of the inverse of the norm, $|\bs{F}_\epsilon\bs{t}|^{-1}$, we establish that
\begin{align}
\dd{(\bs{F}^{\trans}_\epsilon\bs{F}_\epsilon)}{\epsilon}\Big\vert_{\epsilon=0}&=\dd{}{\epsilon}\left[(1-\epsilon)^2\id+\epsilon^2\bs{F}^{\trans}\bs{F}+\epsilon(1-\epsilon)(\bs{F}^{\trans}+\bs{F})\right]_{\epsilon=0},\nonumber\\[4pt]
&=\left[-2(1-\epsilon)\id+2\epsilon\bs{F}^{\trans}\bs{F}+(1-2\epsilon)(\bs{F}^{\trans}+\bs{F})\right]_{\epsilon=0},\nonumber\\[4pt]
&=-2\id+(\bs{F}^{\trans}+\bs{F})=2\mskip+2.5mu\sym(\Grad\bs{u}),
\end{align}
to compute the rate of change of the norm, $|\bs{F}_\epsilon\bs{t}|=\sqrt{\bs{F}^{\trans}_\epsilon\bs{F}_\epsilon\colon\bs{t}\otimes\bs{t}}$,
\begin{equation}
\dd{(\sqrt{\bs{F}^{\trans}_\epsilon\bs{F}_\epsilon\colon\bs{t}\otimes\bs{t}})}{\epsilon}\Big\vert_{\epsilon=0}=\sym(\Grad\bs{u})\colon\bs{t}\otimes\bs{t},
\end{equation}
to arrive at
\begin{equation}\label{eq:rate.inv.norm.tangent}
\dd{(|\bs{F}_\epsilon\bs{t}|^{-1})}{\epsilon}\Big\vert_{\epsilon=0}=-\sym(\Grad\bs{u})\colon\bs{t}\otimes\bs{t}=-\Grad\bs{u}\colon\bs{t}\otimes\bs{t}.
\end{equation}
Thus, from expressions \eqref{eq:rate.tangent.no.preserving} and \eqref{eq:rate.inv.norm.tangent}, we determine the rate of change of the unit tangent which preserves the norm
\begin{align}\label{eq:rate.tangent}
\dd{\bs{t}_\epsilon}{\epsilon}\Big\vert_{\epsilon=0}&=-(\Grad\bs{u}\colon\bs{t}\otimes\bs{t})\bs{t}+(\Grad\bs{u})\bs{t},\nonumber\\[4pt]
&=\left(-\bs{t}\otimes\pards{\bs{u}}{s}+\Grad\bs{u}\right)\bs{t}=-\left(\pards{\bs{u}}{s}\cdot\bs{t}\right)\bs{t}+\pards{\bs{u}}{s}.
\end{align}

From the definition of the unit tangent-normal $\bs{\nu}\coloneqq\bs{t}\times\bs{n}$, we determine the rate of change \eqref{eq:rates.tensors}$_{3}$, using \eqref{eq:rate.tangent} and \eqref{eq:rate.normal}, we obtain the change of rate of the unit tangent-normal
\begin{align}
\dd{\bs{\nu}_\epsilon}{\epsilon}\Big\vert_{\epsilon=0}&=\dd{\bs{t}_\epsilon}{\epsilon}\Big\vert_{\epsilon=0}\times\bs{n}+\bs{t}\times\dd{\bs{n}_\epsilon}{\epsilon}\Big\vert_{\epsilon=0},\nonumber\\[4pt]
&=-\left(\left(\pards{\bs{u}}{s}\cdot\bs{t}\right)\bs{t}-\pards{\bs{u}}{s}\right)\times\bs{n}+(\Grads\bs{u})^{\trans}\bs{n}\times\bs{t}.
\end{align}

With the following
\begin{equation}
(\Grads\bs{u})^{\trans}\bs{n}=\Grads(\bs{u}\cdot\bs{n})-(\Grads\bs{n})^{\trans}\bs{u}=\Grads(\bs{u}\cdot\bs{n})+\bs{L}\bs{P}\bs{u},
\end{equation}
\begin{equation}
\pards{\bs{u}}{s}=\left(\pards{\bs{u}}{s}\cdot\bs{t}\right)\bs{t}+\left(\pards{\bs{u}}{s}\cdot\bs{n}\right)\bs{n}+\left(\pards{\bs{u}}{s}\cdot\bs{\nu}\right)\bs{\nu},
\end{equation}
and
\begin{equation}
\left(\Grads(\bs{u}\cdot\bs{n})+\bs{L}\bs{P}\bs{u}\right)\times\bs{t}=\left(\left(\Grads(\bs{u}\cdot\bs{n})+\bs{L}\bs{P}\bs{u}\right)\cdot\bs{\nu}\right)\bs{n},
\end{equation}
we find the following alternative representations for the rate of change of $\bs{n}_\epsilon$, $\bs{t}_\epsilon$, and $\bs{\nu}_\epsilon$ with respect to $\epsilon$,
\begin{subequations}
\begin{align}
\dd{\bs{n}_\epsilon}{\epsilon}\Big\vert_{\epsilon=0}&=-\Grads(\bs{u}\cdot\bs{n})-\bs{L}\bs{P}\bs{u},\\[4pt]
\dd{\bs{t}_\epsilon}{\epsilon}\Big\vert_{\epsilon=0}&=\left(\pards{\bs{u}}{s}\cdot\bs{n}\right)\bs{n}+\left(\pards{\bs{u}}{s}\cdot\bs{\nu}\right)\bs{\nu},\\[4pt]
\dd{\bs{\nu}_\epsilon}{\epsilon}\Big\vert_{\epsilon=0}&=-\left(\pards{\bs{u}}{s}\cdot\bs{\nu}\right)\bs{t}+\left(\left(\Grads(\bs{u}\cdot\bs{n})+\bs{L}\bs{P}\bs{u}\right)\cdot\bs{\nu}\right)\bs{n}.
\end{align}
\end{subequations}

We now aim at reckoning the rates of change \eqref{eq:rates.tensors}$_{5,6,7}$ of the differential elements of length \eqref{eq:ds}, area \eqref{eq:da}, and volume \eqref{eq:dv}. The rates of change are given as follows. The rate of change of the differential element of length is
\begin{equation}
\dd{|\bs{F}_\epsilon\bs{t}|}{\epsilon}\Big\vert_{\epsilon=0}=\Grad\bs{u}\colon\bs{t}\otimes\bs{t},
\end{equation}
while the rate of change of the differential element of volume is
\begin{align}
\dd{J}{\epsilon}\Big\vert_{\epsilon=0}&=\dd{}{\epsilon}\left((\bs{F}_\epsilon\bs{t}\times\bs{F}_\epsilon\bs{n})\cdot\bs{F}_\epsilon\bs{\nu}\right)\Big\vert_{\epsilon=0}\nonumber\\[4pt]
&=\Grad\bs{u}\colon(\bs{t}\otimes\bs{t}+\bs{n}\otimes\bs{n}+\bs{\nu}\otimes\bs{\nu})=\Div\bs{u},
\end{align}
and, finally, the rate of change of the differential element of area is
\begin{equation}
\dd{(J|\bs{F}^{-\trans}_\epsilon\bs{n}|)}{\epsilon}\Big\vert_{\epsilon=0}=\Divs\bs{u}.
\end{equation}

Finally, we parameterize the curvature tensor $\bs{L}$ as $\bs{L}_\epsilon\coloneqq-\Grads{}_{\!_\epsilon}\bs{n}_\epsilon$ on $\srf_\epsilon$. We can write the gradient on $\srf_\epsilon$ as
\begin{equation}
\Grad\bs{n}_\epsilon\eqqcolon\Grad_{\bs{x}}\bs{n}_\epsilon=(\Grad_{\bs{y}_\epsilon}\bs{n}_\epsilon)\Grad_{\bs{x}}\bs{y}_\epsilon,\quad\text{or}\quad\Grad_{\bs{y}_\epsilon}\bs{n}_\epsilon=\Grad_{\bs{x}}\bs{n}_\epsilon(\Grad_{\bs{x}}\bs{y}_\epsilon)^{-1},
\end{equation}
and keeping the use of the deformation gradient
\begin{equation}
\Grad_{\bs{y}_{\epsilon}}\bs{n}_\epsilon=\Grad_{\bs{x}}\bs{n}_\epsilon(\bs{F}^{-1}_\epsilon).
\end{equation}
Thus, we rewrite the curvature tensor as
\begin{equation}
\bs{L}_\epsilon=-\Grad\bs{n}_\epsilon(\bs{F}^{-1}_\epsilon)(\id-\bs{n}_\epsilon\otimes\bs{n}_\epsilon).
\end{equation}
We can now compute the first variation of the curvature tensor as
\begin{align}
\dd{\bs{L}_\epsilon}{\epsilon}\Big\vert_\epsilon=&\,\Grad\left((\Grads\bs{u})^{\trans}\bs{n}\right)(\id-\bs{n}\otimes\bs{n})+\Grad\bs{n}\mskip+2.5mu\Grad\bs{u}(\id-\bs{n}\otimes\bs{n})\nonumber\\
&-\Grad\bs{n}\left((\Grads\bs{u})^{\trans}\bs{n}\otimes\bs{n}+\bs{n}\otimes(\Grads\bs{u})^{\trans}\bs{n}\right)\nonumber\\
=&\,\Grads\left((\Grads\bs{u})^{\trans}\bs{n}\right)-\bs{L}\mskip+2.5mu\Grads\bs{u}+\bs{L}(\Grads\bs{u})^{\trans}\bs{n}\otimes\bs{n},
\end{align}
where we use the fact that $\bs{n}\otimes(\Grads\bs{u})^{\trans}\bs{n}=(\bs{n}\otimes\bs{n})\Grads\bs{u}$.


\footnotesize

\bibliographystyle{unsrt}

\begin{thebibliography}{10}

\bibitem{Fri93}
E~Fried and ME~Gurtin.
\newblock Continuum theory of thermally induced phase transitions based on an
  order parameter.
\newblock {\em Physica D: Nonlinear Phenomena}, 68(3-4):326--343, 1993.

\bibitem{Fri94}
E~Fried and ME~Gurtin.
\newblock Dynamic solid-solid transitions with phase characterized by an order
  parameter.
\newblock {\em Physica D: Nonlinear Phenomena}, 72(4):287--308, 1994.

\bibitem{Gur96}
ME~Gurtin.
\newblock Generalized {G}inzburg--{L}andau and {C}ahn--{H}illiard equations
  based on a microforce balance.
\newblock {\em Physica D: Nonlinear Phenomena}, 92(3-4):178--192, 1996.

\bibitem{Fos89}
RL~Fosdick and EG~Virga.
\newblock A variational proof of the stress theorem of cauchy.
\newblock {\em Archive for Rational Mechanics and Analysis}, 105(2):95--103,
  1989.

\bibitem{Fos16}
R~Fosdick.
\newblock A generalized continuum theory with internal corner and surface
  contact interactions.
\newblock {\em Continuum Mechanics and Thermodynamics}, 28(1-2):275, 2016.

\bibitem{Swi77}
J~Swift and PC~Hohenberg.
\newblock Hydrodynamic fluctuations at the convective instability.
\newblock {\em Physical Review A}, 15(1):319, 1977.

\bibitem{Bra75}
SA~Brazovski\v{\i}.
\newblock Phase transition of an isotropic system to a nonuniform state.
\newblock {\em Soviet Journal of Experimental and Theoretical Physics}, 41:85,
  1975.

\bibitem{Esp17}
LFR Espath, AF~Sarmiento, Lisandro Dalcin, and VM~Calo.
\newblock On the thermodynamics of the swift--hohenberg theory.
\newblock {\em Continuum Mechanics and Thermodynamics}, 29(6):1335--1345, 2017.

\bibitem{Esp20}
L~Espath, V~Calo, and E~Fried.
\newblock Generalized swift–hohenberg and phase-field-crystal equations based
  on a second-gradient phase-field theory.
\newblock {\em submitted}, --(--):--, 2020.

\bibitem{Tru04}
C~Truesdell and W~Noll.
\newblock The non-linear field theories of mechanics.
\newblock In {\em The non-linear field theories of mechanics}, pages 1--579.
  Springer, 2004.

\bibitem{Gur02}
ME~Gurtin.
\newblock A gradient theory of single-crystal viscoplasticity that accounts for
  geometrically necessary dislocations.
\newblock {\em Journal of the Mechanics and Physics of Solids}, 50(1):5--32,
  2002.

\bibitem{Col63}
BD~Coleman and W~Noll.
\newblock The thermodynamics of elastic materials with heat conduction and
  viscosity.
\newblock {\em Archive for Rational Mechanics and Analysis}, 13(1):167--178,
  1963.

\bibitem{Fri06b}
E~Fried.
\newblock On the relationship between supplemental balances in two theories for
  pure interface motion.
\newblock {\em SIAM Journal on Applied Mathematics}, 66(4):1130--1149, 2006.

\bibitem{Fri06a}
E~Fried and ME~Gurtin.
\newblock Tractions, balances, and boundary conditions for nonsimple materials
  with application to liquid flow at small-length scales.
\newblock {\em Archive for Rational Mechanics and Analysis}, 182(3):513--554,
  2006.

\bibitem{Fri07}
E.~Fried and M.~E. Gurtin.
\newblock Thermomechanics of the interface between a body and its environment.
\newblock {\em Continuum Mechanics and Thermodynamics}, 19(5):253--271, 2007.

\bibitem{Dud19}
F.~P. Duda, A.~Sarmiento, and E.~Fried.
\newblock Phase fields, constraints, and the {C}ahn--{H}illiard equation.
\newblock {\em Submitted}, 2019.

\bibitem{Gur10}
ME~Gurtin, E~Fried, and L~Anand.
\newblock {\em The mechanics and thermodynamics of continua}.
\newblock Cambridge University Press, 2010.

\bibitem{Ste98}
GW~Stewart.
\newblock {\em Matrix Algorithms: Basic Decompositions}.
\newblock SIAM, Society for industrial and applied mathematics, 1998.

\end{thebibliography}

\end{document}